# Broadband nonvolatile electrically programmable silicon photonic switches

Rui Chen,[1] Zhuoran Fang,[1] Johannes E. Fröch,[1] Peipeng Xu,[2] Jiajiu Zheng, [1]* Arka Majumdar[1,3]*

[1]Department of Electrical and Computer Engineering, University of Washington, Seattle, WA 98195, USA

[2]Faculty of Electrical Engineering and Computer Science, Key Laboratory of Photoelectric Materials and Devices of Zhejiang Province, Ningbo University, Ningbo, 315211, China

[3]Department of Physics, University of Washington, Seattle, WA 98195, USA

*jjzno1@gmail.com, arka@uw.edu

## Abstract

Programmable photonic integrated circuits (PICs) have recently gained significant interest due to their potential in creating next-generation technologies ranging from artificial neural networks and microwave photonics to quantum information processing. The fundamental building block of such programmable PICs is a tunable 2 × 2 switch, traditionally controlled by the thermo-optic or free-carrier dispersion. Yet, these implementations are power-hungry, volatile, and have a large footprint (typically > 100 μm). Therefore, a truly 'set-and-forget' type 2 × 2 switch with zero static power consumption is highly desirable for large-scale PICs. Here, we report a broadband nonvolatile electrically programmable 2 × 2 silicon photonic switch based on the phase-change material $Ge_2Sb_2Te_5$. The directional coupler switch exhibits a compact coupling length (64 μm), small insertion loss (<2 dB), and minimal crosstalk (<-8 dB) across the entire telecommunication C-band while maintaining a record-high endurance of over 2,800 switching cycles. This demonstrated switch constitutes a critical component for realizing future generic programmable silicon photonic systems.

# Introduction

Photonic integrated circuits (PICs) typically have been application-specific (*1*), i.e., each fabricated chip serving only one particular function. In contrast, recent advancements in silicon photonics urgently call for universally programmable PICs(*2*), adaptable to a wide range of tasks, including optical neural networks(*3*), quantum information processing(*4*), and light detection and ranging(*5*). The fundamental building block constituting such programmable PICs is an electrically controlled tunable 2 × 2 switch that diverts light to one of two ports. In silicon photonics, the operation of such elements is commonly based on either the thermo-optic(*6*) or free-carrier effect(*7*). However, their functions are plagued by large footprints (typically > 100 μm) and/ or low energy efficiency due to large static power consumption. Although ring resonators can significantly reduce the footprint (down to the order of tens of μm), the operational optical bandwidth becomes limited to typically less than 1 nm(*8*). Moreover, their thermal instability requires an even larger static power consumption, as a constant feedback signal is needed to lock the resonance(*9*). Emerging technologies such as microelectromechanical systems (MEMS) (*10*), electro-optic polymer(*11*), and integrated lithium niobate(*12*) can potentially provide either a smaller footprint or lower programming energy. Still, they are all based on a volatile effect with limited CMOS compatibility. Phase-change materials (PCMs) provide an attractive solution towards compact and energy-efficient 2 × 2 switches with zero static power(*13–16*), thanks to a nonvolatile phase transition(*17*), large refractive index contrast ($\Delta n \geq 1$)(*18*), and CMOS compatibility(*15*).

Pulsed laser induced switching of PCMs for all-optical applications, such as optical memories(*19–23*) and optical computing(*24–27*), has been studied extensively. However, this approach usually involves complex optical alignment, which precludes simultaneous multi-device control and is generally unsuitable for large-scale applications. Recently the phase transition has been actuated on-chip electrically via doped silicon(*28–30*) and/or ITO heaters(*31–33*), achieving reversible tuning and a large cyclability of ~500 cycles(*28*). Yet, thus far, no broadband, multi-port, electrically programmable device based on $Ge_2Sb_2Te_5$ (GST) has been demonstrated. While such functionality is ultimately required for large-scale nonvolatile photonic architectures, the inherent high absorptive loss of crystalline GST (c-GST) (*34*) becomes a severe issue for most designs (See Supplementary, Section 2.1). Existing electrically tunable GST devices are either designed only for 1 × 1 switches(*28*, *32*) or rely on ring resonators(*28*, *35*) to circumvent the loss issue. While the latter is suitable for narrowband operation, a broadband design is generally preferred to access a wider spectral range.

Here we report a broadband nonvolatile electrically programmable 2 × 2 silicon photonic switch based on the technologically mature PCM GST. The directional coupler-based switch features a compact footprint of only 64 μm coupling length with less than 2 dB insertion loss and -8 dB crosstalk in both bar and cross states across the entire telecommunication C-band. The fabricated device displays a record-high endurance for operation over 2,800 switching cycles, *i.e.*, over 5,600 switching events. The fabrication of the switch is fully compatible with standard CMOS foundry processes, promising excellent scalability. Therefore, our work represents a crucial step towards realizing compact and energy-efficient programmable silicon PICs.

# Results

Our electrically tunable 2 × 2 switch leverages an asymmetrical three-waveguide directional coupler structure(*13*, *36*) fabricated in a Silicon on Insulator (SOI) platform, as shown in *Fig. 1a*. It consists of two Si transmission waveguides with a coupling region separated by a truncated transition waveguide. To circumvent the high absorption loss of c-GST, only the transition waveguide is loaded with a 20 nm-thick GST, encapsulated by 40 nm $Al_2O_3$ to prevent GST oxidation. By design, the phase-matching condition is satisfied only when the GST is switched to the amorphous state (a-GST), which allows light to couple into the transition waveguide and further to the cross port. However, if the GST is in the crystalline state, light barely couples into the transition waveguide due to a large index mismatch, thus bypassing the high loss of c-GST. The gap between the waveguides is designed to achieve a good tradeoff between insertion loss and device footprint (Supplementary, Section 2). While similar designs have been reported before(*13*, *36*), no electrical control has been demonstrated, primarily due to the lack of suitable heater designs for large-volume GST switching. A PIN heating element implements the electrical control over the phase switching, which is realized by selectively doping the adjacent regions to the waveguide. We note that this concept directly extends our prior works, where we implemented this concept(*28*) for a 1 × 1 switch. However, in contrast to that work, the volume of the GST being electrically switched experimentally is ~0.4 $\mu m^3$, around ten times larger(*28*, *31*–*33*). Therefore, prior to fabrication, we optimized the design and operation of the PIN microheater and device transmission using finite element simulations(*28*, *37*), considering the electrical and heat transfer transient performance of the device (Supplementary, Section 3).

The fabricated device's optical and scanning electron microscope (SEM) images are shown in *Fig. 1b* and *1c*, respectively. In detail, the phase of the PCM is switched by an on-chip PIN heater, defined by doping the 100-nm-thick silicon slab via ion implantation while leaving the three waveguides intrinsic(*28*) to reduce the insertion loss (see *Methods* for fabrication details). The doped regions and metal pads are designed to maintain a distance of 200 nm and 1 μm, respectively, from the directional coupler to reduce excess loss from the free carriers.

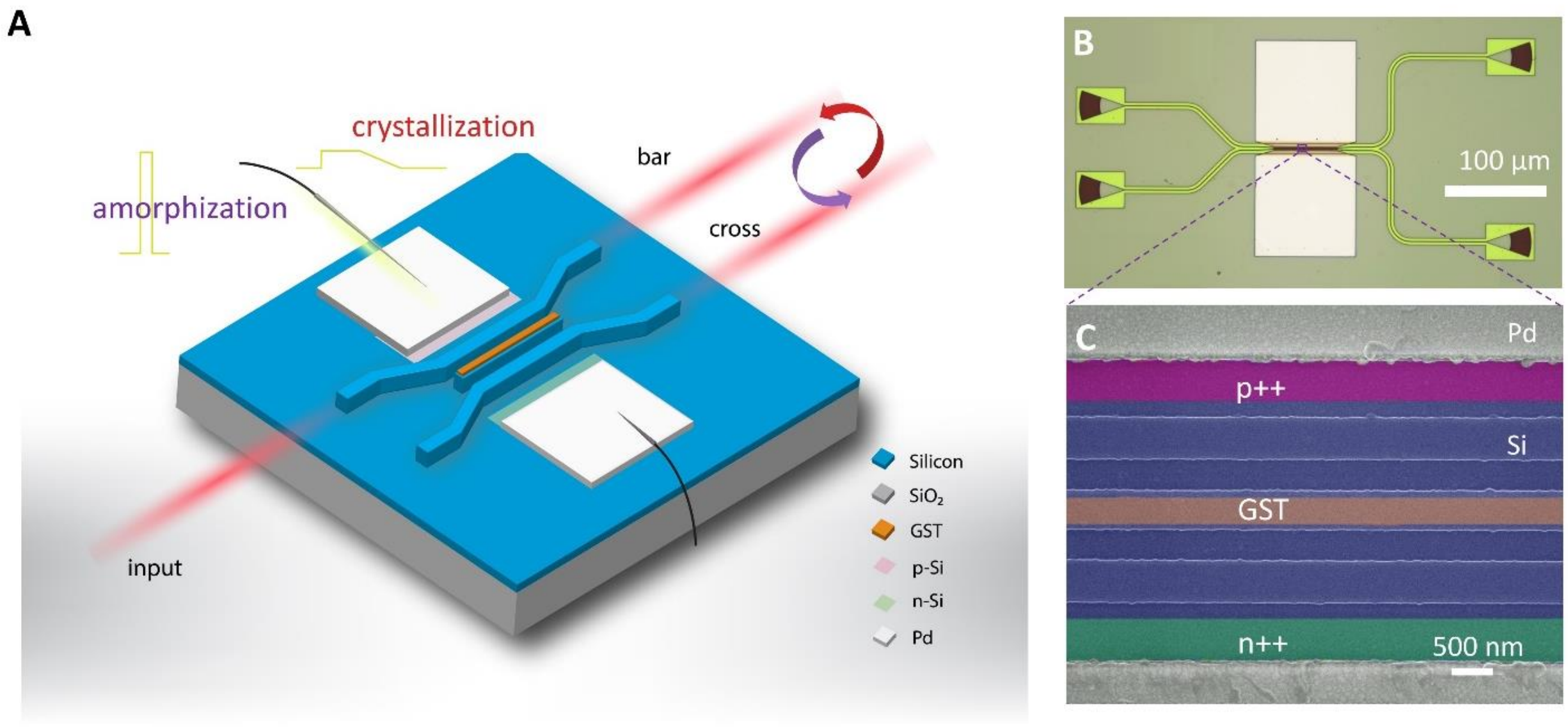

Figure 1. **Nonvolatile electrically controllable 2 × 2 broadband photonic switch.** (a) Device schematic. (b) The optical micrograph and (c) SEM images of the device. In the SEM image, the GST thin film, p++, and n++ regions are indicated by false colors.

The reversible operation of the 2 × 2 switch is demonstrated by measuring the transmission of a tunable laser through the bar and cross port at two different phases of GST, summarized in *Fig. 2a*. Initially, the GST is prepared in the crystalline state (c-GST), for which the light passes through the bar port with an extinction ratio over 10 dB (*Fig. 2a(i))*. After applying a short, high amplitude pulse (13.6 V, 8 ns leading/falling edge, 200 ns pulse width with 380 nJ pulse energy) to induce amorphization (a-GST), the transmission spectrum flips, resulting in a high transmission through the cross port with over 10 dB extinction ratio (*Fig. 2a(ii)*). Subsequently, we applied a longer but lower amplitude pulse (3.2 V, 8 ns leading edge, 50 μs pulse width with 6.83 μJ pulse energy) to trigger crystallization, returning the switch to the original high bar transmission state (*Fig. 2a(iii)*). Crosstalk less than -8 dB is observed across the entire telecommunication C-band (1530 to 1565 nm) in cross and bar states. We estimate the switching energy density to be 0.95/1.63 fJ/nm$^3$ for amorphization/crystallization. This is comparable with (*28*), where an energy density of 0.2/1.95 fJ/nm$^3$ for the phase transition was demonstrated. The cycle-to-cycle reproducibility was verified over three consecutive switching cycles, as shown in *Fig. 2b*, where transmission spectra for a-GST and c-GST at bar port overlap nearly perfectly. We then established the long-term retention of the device states by performing several measurements over a time period of one month, as shown in *Fig. 2c*. A good match between the transmission spectra for devices in the a-GST configuration after putting the device in an ambient environment for 132 hours and 720 hours indicates that the switching is indeed nonvolatile. Slight variations among the spectra are likely due to minor differences in the optical alignment during the different measurements.

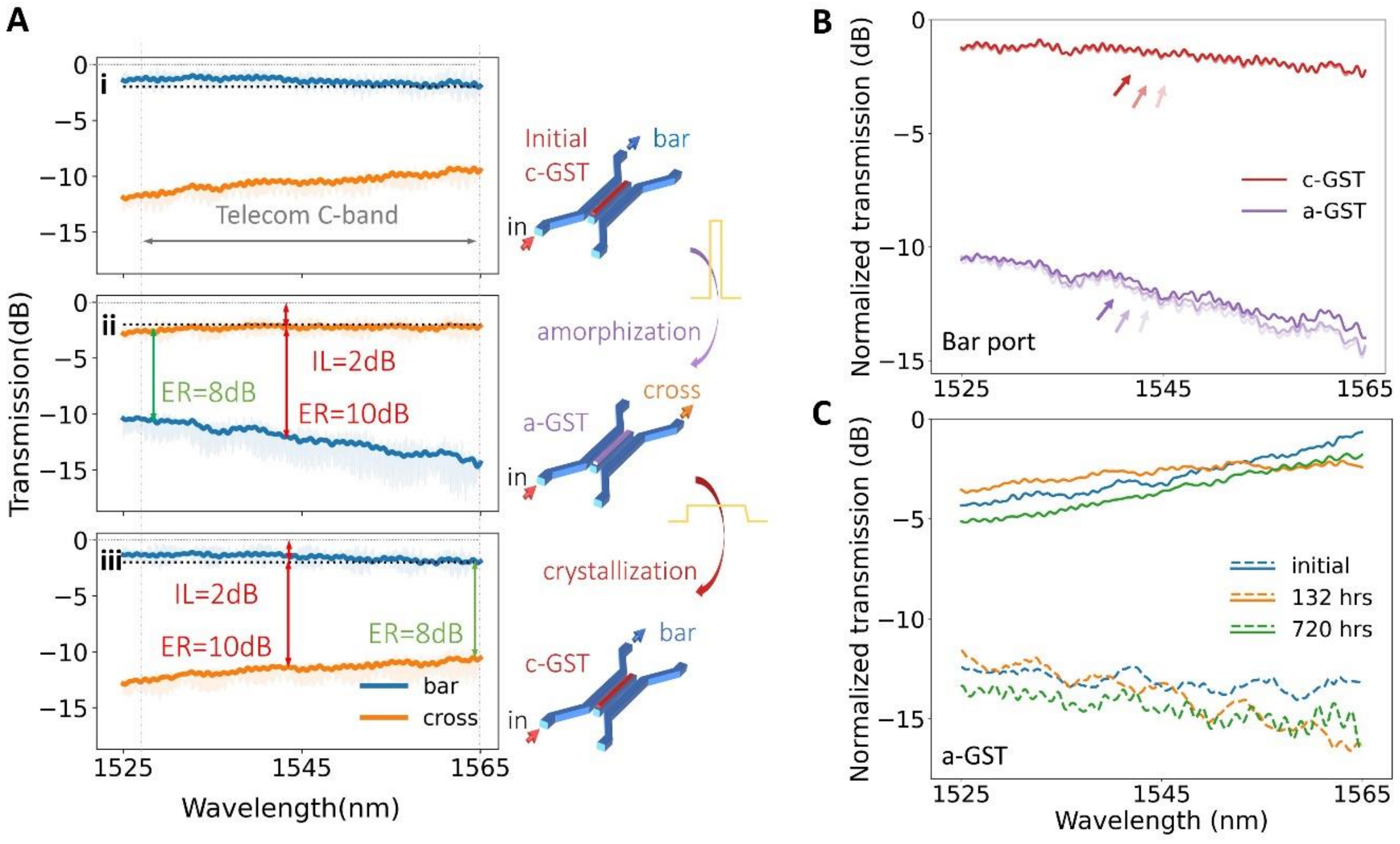

Figure 2. **Characterizations of the 2×2 directional coupler switch.** (a) Transmission spectrum of the device (i) initially, (ii) after applying an amorphization pulse (iii) after applying a crystallization pulse. The shaded areas are the measured raw spectra, and the solid lines are spectra after a low pass filter. IL and ER are insertion loss and extinction ratio, respectively. (b) Transmission spectra measured over three cycles at bar port. Each cycle is indicated with different color saturations. (c) Device functionality over an extended period of one month with a-GST. The solid and dashed lines correspond to the bar and cross port transmission.

Finally, we demonstrate reversible operation over 5,750 switching events without significant performance degradation, presented in *Fig. 3* (additional measurement results are shown in Supplementary, Section 4.1), thanks to the optimized PIN heater design. We note that due to limitations in our measurement setup (See Methods for measurement details), the cross and bar ports are measured consecutively, as shown in *Fig. 3(top)* and *(bottom)*, respectively. The two distinct transmission levels arising from switching between the crystalline and amorphous state are indicated by the shaded regions, whereas transmission contrasts of 8 dB and 11 dB are obtained for the cross and bar ports, respectively. The transmission contrast of the cross port reduces from 8 dB to 6 dB at the 2,750$^{th}$ event, while the contrast of the bar port remains around 11 dB. We attribute this reduction in the transmission contrast of the cross port to thermal reflowing and material degradation of GST after multiple cycles (Supplementary, Section 6). The performance degradation is less pronounced for the transmission through the bar port because even as the GST shape changes due to thermal reflowing, a significant index mismatch still exists between the transition waveguide and input waveguide (See Figure S7(E), Supplementary). We note that some data points are removed around the 600$^{th}$, 800$^{th}$, and 1,900$^{th}$ events because the same amorphization condition failed to trigger the phase change, and larger voltages of 13.8, 14.1, and 14.4V were used after that, respectively. This could result from drifting contacts of electrical probes (Supplementary, Section 4, *Fig S10(c)*) and/ or GST degradation. The contact drift issue could be solved by wire-bonding the chip to a carrier.

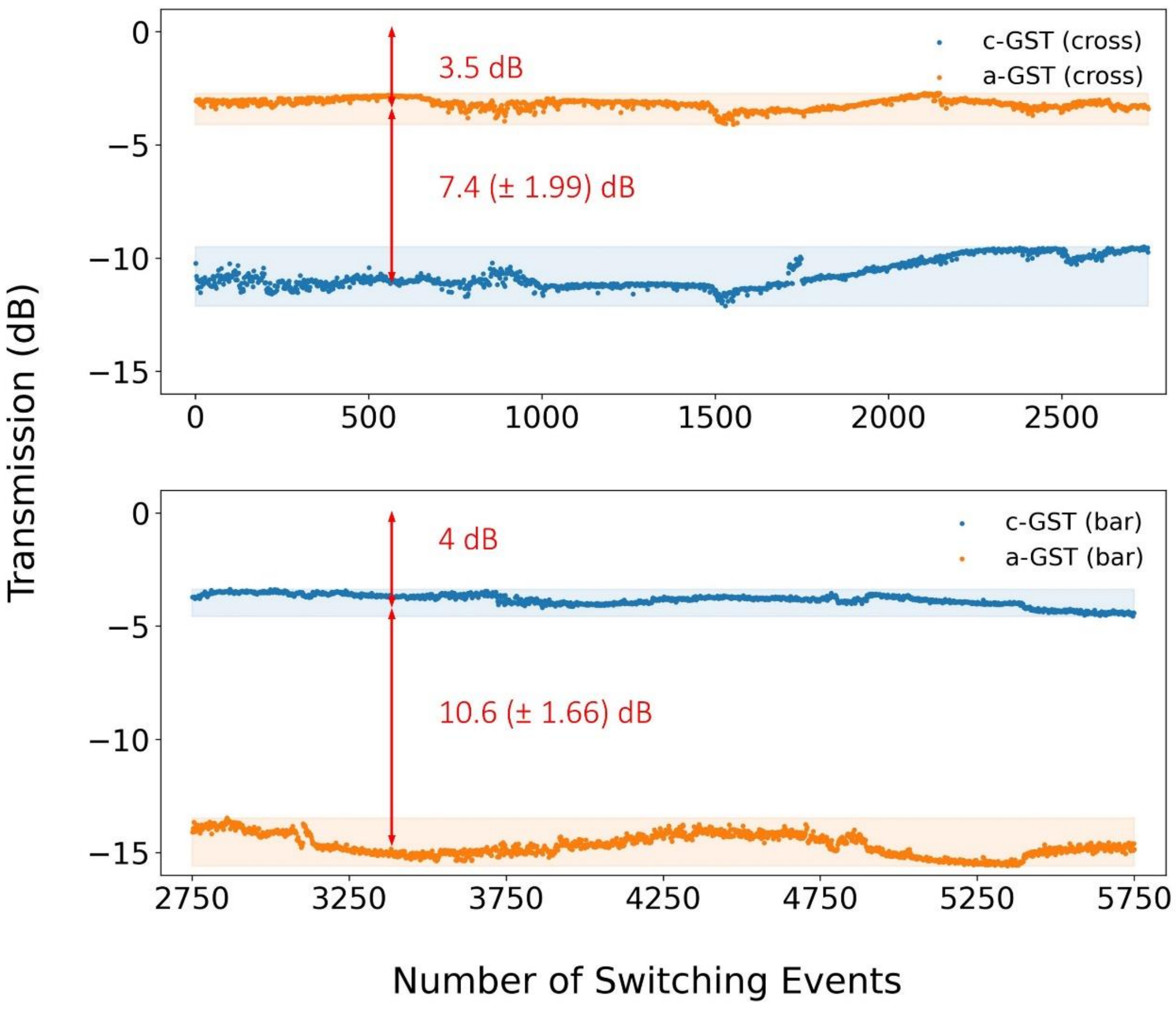

Figure. 3 **Endurance test of the 2×2 switch.** (Top) Transmission measured at the cross port. (Bottom) Transmission measured at the bar port. The blue/orange dots represent the measured transmission at 1550 nm after sending in a crystallization/amorphization pulse (with the same pulse condition in Fig. 2). The shaded regions correspond to variations of the two distinct transmission levels. A larger insertion loss of 3.5 (4.0) dB at cross (bar) port is attributed to the material degradation and thermal reflowing of the GST thin film, see Discussion Section.

In addition to the 2 × 2 switch, we also demonstrated electrically tunable 1 × 2 and waveguide-based 1 × 1 switches based on the GST-on-SOI platform (Supplementary Materials). The 1 × 2 switch consists of two mutually coupled waveguides, one of which is loaded with 20 nm GST thin film. The two waveguides are designed in a similar fashion as in the 2 × 2 switch that the phase-matching condition is satisfied when GST is in the amorphous state, while a large mismatch happens when it is switched into the crystalline state. Light from the bare silicon waveguide can then efficiently couple into the GST-loaded waveguide for a-GST, but remains in the bar waveguide for c-GST. This switch shows an extinction ratio of 9 dB over 15 nm wavelength range in both amorphous and crystalline states and an excellent endurance of more than 4,000 cycles (Supplementary, Section 4.2). The waveguide-based 1 × 1 switch is a 10-μm long waveguide loaded with 20 nm thick GST thin film. The switch exhibits a large extinction ratio of 15 dB, the capability of multi-level switching with ten distinct transmission levels, and excellent endurance of more than 1000 switching cycles (Supplementary, Section 4.3). Potential applications of these switches are discussed in Supplementary Section 7, where nonvolatile PICs for energy-efficient optical routing, OPGA, and optical computing are proposed.

# Discussion

Table 1 compares our device to other reported tunable 2 × 2 switches over different performance metrics. For switches based on a volatile effect, the energy consumption equals the energy per second. For devices based on nonvolatile effects, the energy consumption is defined as the energy per switching event. This comparison is reasonable because programmable PICs, such as for optical information processing(*3*, *24*, *27*), usually work in the low electrical frequency, *i.e.*, the switches are not turned on and off constantly in a short period. Hence, volatile switches impose a constant drain in the energy over one second, while the nonvolatile ones do not consume energy after the initial switching event. From *Table 1*, our 2 × 2 switch shows several orders of magnitude reduction in device footprint and energy consumption compared to the volatile schemes except for the MEMS, which is inherently challenging to fabricate. Our work also compares favorably to previously reported PCM-based photonic switches(*13*, *28*, *35*, *38*, *39*) in terms of the largest reported bandwidth (> 30 nm), electrical tunability, large extinction ratio in both cross and bar states (> 10 dB), and the highest reported endurance (> 2,800 cycles).

We further demonstrate the thermal stability advantage of our broadband switches over the narrowband designs by a thermal stability test. It is well-known that thermal feedback control is required for the stable operation of micro-ring resonators due to their sensitivity to thermal fluctuation(*9*). We performed a thermal stability test on our broadband switch (Supplementary, Section 5) by varying the temperature from 25°C to 35°C, and no significant spectrum drift was observed. Such invariant to thermal fluctuations renders our switches even more energy efficient as no extra energy is needed for temperature control.

The relatively high insertion loss of our 2 × 2 switches in the experiment (~ 2 dB at 1550 nm) compared to the simulation (< 1 dB, see Supplementary, Section 2.3) can be attributed to several factors. First, fabrication imperfections such as variations in etch depth and waveguide, and GST width (Supplementary, Section 6, Figure S15) could all lead to an increase in device insertion loss (Supplementary, Section 2.4). Hence, we believe the insertion loss can be reduced to < 1 dB based on our simulations with optimized fabrication conditions. Second, a tradeoff between the insertion loss in the two states can be observed in *Fig. S4* and *S5* (Supplementary, Section S2) due to finite loss of amorphous GST. This can be resolved by replacing GST with recently reported low-loss PCMs such as $Sb_2S_3$ and $Sb_2Se_3$(*40–43*). However, we note that the choice of GST is preferred for compatibility with existing high-volume manufacturing since this material has been extensively studied in electronic memories and is technologically mature(*17*, *44*).

Although our 2 × 2 nonvolatile switch displayed the highest reported cyclability among electrically controlled PCM-based photonic switches, it is still significantly lower than the records obtained for electronic memories (~$10^{12}$ cycles, potentially scale up to $10^{15}$cycles) (*17*, *44*). At this point, the cyclability of our device is mainly limited by thermal reflowing and material ablation due to the large GST volume (Supplementary, Section 6). To improve the cyclability further, instead of using a long (64 µm) GST thin film, one can leverage sub-wavelength grating structure, where periodic GST nano-disks or nano-strips with much smaller length (down to 100 nm) can significantly reduce the volume of each GST cell, thus ease the thermal reflow issue (*27*, *35*). A thicker $Al_2O_3$ capping layer could also be used to improve the encapsulation, for example, 218 nm $Al_2O_3$ capping layer is used in Ref. (*27*).

**Table 1. Tunable optical 1 × 2 or 2 × 2 switches performance comparison**

| Ref | Structure | Mechanism | ER (dB) | IL (dB) | Energy per second (nJ/s) | Footprint in length (µm) | Optical BW (nm) | Nonvolatile | Tuning method and cycles |
|---|---|---|---|---|---|---|---|---|---|
| (*45*) | MZI | Thermal | > 20.0 | 0.5 | $1.3 \times 10^7$ | > 200 | > 70 | No | Electrical |
| (*7*) | MZI | Free carrier | 4.5 | 10.0 | $2.7 \times 10^9$ | $1.3 \times 10^4$ | N. R. | No | Electrical |
| (*12*) | MZI | EO LN | 40.0 | 2.5 | $1.9 \times 10^7$ | $5.0 \times 10^4$ | > 40 | No | Electrical |
| (*46*) | MZI | EO polymer | > 25.0 | 8.2 | $8.5 \times 10^6$ | $1.5 \times 10^4$ | N. R. | No | Electrical |
| (*8*) | MDR | Thermal | 7.0 | 0.8 / 9.0 * | $1.5 \times 10^7$ | 12.8 | < 0.5 | No | Electrical |
| (*47*) | MRR | Free carrier | > 20.0 | ~ 0 | $9.7 \times 10^6$ | 35 | < 0.1 | No | Electrical |
| (*48*) | MRR | MEMS | > 20.0 | 2.0 / 0.1 * | 600 | 4 | < 2 | No | Electrical |
| (*39*) | MRR | PCM | > 5.0 | 5.1 / 4.3 * | 0.19 (17.1) ‡ | 60 | < 1 | Yes | Optical, 1,000 cycles |
| (*35*) | MRR | PCM | 14 | 0.75 / 0.46 * | 0.25 (11) ‡ | 25 | < 1 | Yes | Optical |
| (*13*) | DC | PCM | > 10 | 1.0 | N. R. | 45 | > 30 | Yes | RTA |
| (*38*) | MMI | PCM | 8.0 | 0.5 | 14 ($9.5 \times 10^5$) ‡ | 43 | N. R. | Yes | Optical |
| (*30*) | MZI | PCM | 6.5 (cross) /15.0 (bar) | > 0.3 | 176 ($3.8 \times 10^3$) ‡ | > 100 † | > 15 † | Yes | Electric, >125 cycles |
| This work | DC | PCM | 10.0 | 2.0 | 380 ($6.8 \times 10^3$) ‡ | 50 | > 30 | Yes | Electric, >2,800 cycles |

(ER: extinction ratio, IL: insertion loss, BW: bandwidth, MZI: Mach-Zehnder Interferometer, MR(D)R: Micro-ring(disk) resonator, DC: directional coupler, EO: Electro-optic effect, LN: $LiNbO_3$, N.R.: not reported)

Notes: For devices based on nonvolatile effects, energy consumption is the energy per switching event, assuming one second per switching event. * For drop and through ports, respectively. † Estimated from figures. ‡ Energy per switching event for amorphization (crystallization).

In summary, we have demonstrated a broadband nonvolatile electrically controllable 2 × 2 silicon photonic switch. Driven by an on-chip PIN microheater, the 2 × 2 switch can be reversibly and reliably switched for more than 2,800 cycles without significant performance degradation. The cross and bar state maintain small crosstalk (< -8 dB) and insertion loss (< 2dB) across the entire telecommunication C-band. The switch is compatible with CMOS manufacturing processes and has a compact footprint of 64 μm coupling length. The insertion loss of the device can be further reduced by more optimized fabrication and by replacing GST with low-loss PCMs(*40–43*). This work constitutes a crucial element for realizing energy-efficient and compact universal programmable PICs for on-chip nonvolatile light routers, optical programmable gate arrays, neuromorphic computing, and quantum information processing.

# Materials and Methods

## Fabrication process

Our electrically controllable nonvolatile switches were fabricated on a commercial SOI Wafer with 220nm-thick Silicon on 3μm-thick $SiO_2$ (SOITECH). The rib waveguides and grating couplers were defined with electron-beam lithography (EBL, JEOL JBX-6300FS) using a positive tone E-beam resist (200-nm-thick ZEP-520A) and partially etched by 120 nm in a fluorine-based inductively coupled plasma etcher (ICP, Oxford PlasmaLab 100 ICP-18) using a mixed gas of $SF_6$ and $C_4F_8$. The etching rate was calibrated each time before the etching to ensure a correct etch depth. The doping regions were defined by two additional EBL steps with 600-nm-thick poly (methyl methacrylate) (PMMA) resist and implanted by boron (phosphorus) ions for p++(n++) with a dosage of $2 \times 10^{15}$ ions per $cm^2$ and ion energy of 14 keV (40 keV). A tilt angle of 7° while conducting ion implantation was used to misalign with the silicon lattice and thus achieve uniform deep doping. To activate the dopants, the chips were annealed at 950 °C for 10 min (Expertech CRT200 Anneal Furnace). Before metal contact deposition, to ensure ideal Ohmic contact, the surface native oxide was removed by immersing the chips in 10:1 buffered oxide etchant (BOE) for 10 seconds. The metal contacts were then immediately patterned by a fourth EBL step using PMMA and formed by electron-beam evaporation (CHA SEC-600) and lift-off of Ti/Pd (5 nm/180 nm) layers. After a fifth EBL defining the GST window, a 20-nm GST thin film was deposited using a GST target (AJA International) in a magnetron sputtering system (Lesker Lab 18), followed by a lift-off process. The deposition rate was calibrated on a silicon chip using Ellipsometer (Woollam alpha-SE). The GST is then encapsulated by 40-nm-thick $Al_2O_3$ through ALD (Oxford Plasmalab 80PLUS OpAL ALD) at 150 °C. To ensure good contact between the electric probe and metal pads while measuring, the $Al_2O_3$ on the metal contacts were removed by defining a window using a sixth EBL with 600nm PMMA, then etching in a chlorine-based inductively coupled plasma etcher (ICP-RIE, Oxford PlasmaLab 100 ICP-18). Finally, rapid thermal annealing (RTA) at 200 °C for 10 min was conducted to completely crystallize the GST.

## Optical simulation

The refractive index data for GST were measured and fitted by an ellipsometer (Woollam M-2000)(*13*). The nonvolatile switches were designed(verified) by a commercial photonic simulation software package Lumerical MODE(FDTD).

**Heat transfer simulation**

The heat transfer performance of the silicon PIN microheater was simulated with a commercial Multiphysics simulation software COMSOL Multiphysics(*28*, *37*). In the simulation, a heat transfer model is coupled with a semiconductor model to simulate the transient time performance of the microheater and to determine the phase change conditions.

A detailed description of the simulations can be found in Supplementary Material, Section 2 and 3.

**Measurement and characterization**

The photonic switches were measured with a vertical fiber-coupling setup using a coupling angle of 25° (Supplementary, Section 1). The stage temperature was fixed at 26 °C controlled by a thermoelectric controller (TEC, TE Technology TC-720). A tunable continuous-wave laser (Santec TSL-510) provided the input light. The polarization was controlled by a manual fiber polarization controller (Thorlabs FPC526) to match the TE mode of the rib waveguides. A low-noise power meter (Keysight 81634B) was used to measure the static optical transmission from the output grating couplers. The transmission spectra of the devices were obtained by normalizing to the closest reference waveguide spectra. For the I–V characterization and on-chip electrical switching, electrical pulses were applied to the on-chip metal contacts via a pair of electrical probes controlled by two probe positioners (Cascade Microtech DPP105-M-AI-S). In particular, the I-V curve measurement was performed using a source meter (Keithley 2450), which is later used to estimate the power of the applied pulses. The crystallization and amorphization pulses were generated from a pulse function arbitrary generator (Keysight 81160A). The tunable laser, power meter, thermal controller, source meter, and pulse function arbitrary generator are controlled by a laptop with a National Instrument interface.

## References

1. X. Chen, M. M. Milosevic, S. Stanković, S. Reynolds, T. D. Bucio, K. Li, D. J. Thomson, F. Gardes, G. T. Reed, The Emergence of Silicon Photonics as a Flexible Technology Platform. *Proceedings of the IEEE*. **106**, 2101–2116 (2018).

2. W. Bogaerts, D. Pérez, J. Capmany, D. A. B. Miller, J. Poon, D. Englund, F. Morichetti, A. Melloni, Programmable photonic circuits. *Nature*. **586**, 207–216 (2020).

3. Y. Shen, N. C. Harris, S. Skirlo, M. Prabhu, T. Baehr-Jones, M. Hochberg, X. Sun, S. Zhao, H. Larochelle, D. Englund, M. Soljačić, Deep learning with coherent nanophotonic circuits. *Nature Photon*. **11**, 441–446 (2017).

4. J. M. Arrazola, V. Bergholm, K. Brádler, T. R. Bromley, M. J. Collins, I. Dhand, A. Fumagalli, T. Gerrits, A. Goussev, L. G. Helt, J. Hundal, T. Isacsson, R. B. Israel, J. Izaac, S. Jahangiri, R. Janik, N. Killoran, S. P. Kumar, J. Lavoie, A. E. Lita, D. H. Mahler, M. Menotti, B. Morrison, S. W. Nam, L. Neuhaus, H. Y. Qi, N. Quesada, A. Repingon, K. K. Sabapathy, M. Schuld, D. Su, J. Swinarton, A. Száva, K. Tan, P. Tan, V. D. Vaidya, Z. Vernon, Z. Zabaneh, Y. Zhang, Quantum circuits with many photons on a programmable nanophotonic chip. *Nature*. **591**, 54–60 (2021).

5. C. Rogers, A. Y. Piggott, D. J. Thomson, R. F. Wiser, I. E. Opris, S. A. Fortune, A. J. Compston, A. Gondarenko, F. Meng, X. Chen, G. T. Reed, R. Nicolaescu, A universal 3D imaging sensor on a silicon photonics platform. *Nature*. **590**, 256–261 (2021).

6. D. Pérez, I. Gasulla, L. Crudgington, D. J. Thomson, A. Z. Khokhar, K. Li, W. Cao, G. Z. Mashanovich, J. Capmany, Multipurpose silicon photonics signal processor core. *Nat Commun*. **8**, 636 (2017).

7. L. Liao, D. Samara-Rubio, M. Morse, A. Liu, D. Hodge, D. Rubin, U. D. Keil, T. Franck, High speed silicon Mach-Zehnder modulator. *Opt. Express, OE*. **13**, 3129–3135 (2005).

8. W. Zhang, J. Yao, Photonic integrated field-programmable disk array signal processor. *Nat Commun*. **11**, 406 (2020).

9. A. H. Atabaki, S. Moazeni, F. Pavanello, H. Gevorgyan, J. Notaros, L. Alloatti, M. T. Wade, C. Sun, S. A. Kruger, H. Meng, K. Al Qubaisi, I. Wang, B. Zhang, A. Khilo, C. V. Baiocco, M. A. Popović, V. M. Stojanović, R. J. Ram, Integrating photonics with silicon nanoelectronics for the next generation of systems on a chip. *Nature*. **556**, 349–354 (2018).

10. C. Errando-Herranz, A. Y. Takabayashi, P. Edinger, H. Sattari, K. B. Gylfason, N. Quack, MEMS for Photonic Integrated Circuits. *IEEE Journal of Selected Topics in Quantum Electronics*. **26**, 1–16 (2020).

11. S. Koeber, R. Palmer, M. Lauermann, W. Heni, D. L. Elder, D. Korn, M. Woessner, L. Alloatti, S. Koenig, P. C. Schindler, H. Yu, W. Bogaerts, L. R. Dalton, W. Freude, J. Leuthold, C. Koos, Femtojoule electro-optic modulation using a silicon–organic hybrid device. *Light: Science & Applications*. **4**, e255–e255 (2015).

12. M. He, M. Xu, Y. Ren, J. Jian, Z. Ruan, Y. Xu, S. Gao, S. Sun, X. Wen, L. Zhou, L. Liu, C. Guo, H. Chen, S. Yu, L. Liu, X. Cai, High-performance hybrid silicon and lithium niobate Mach–Zehnder modulators for 100 Gbit s −1 and beyond. *Nature Photonics*. **13**, 359–364 (2019).

13. P. Xu, J. Zheng, J. K. Doylend, A. Majumdar, Low-Loss and Broadband Nonvolatile Phase-Change Directional Coupler Switches. *ACS Photonics*. **6**, 553–557 (2019).

14. M. Wuttig, H. Bhaskaran, T. Taubner, Phase-change materials for nonvolatile photonic applications. *Nature Photon*. **11**, 465–476 (2017).

15. Z. Fang, R. Chen, J. Zheng, A. Majumdar, Nonvolatile reconfigurable silicon photonics based on phase-change materials. *IEEE Journal of Selected Topics in Quantum Electronics*, 1–1 (2021).

16. S. Abdollahramezani, O. Hemmatyar, H. Taghinejad, A. Krasnok, Y. Kiarashinejad, M. Zandehshahvar, A. Alù, A. Adibi, Tunable nanophotonics enabled by chalcogenide phase-change materials. *Nanophotonics*. **9**, 1189–1241 (2020).

17. S. Raoux, F. Xiong, M. Wuttig, E. Pop, Phase change materials and phase change memory. *MRS Bull.* **39**, 703–710 (2014).

18. K. Shportko, S. Kremers, M. Woda, D. Lencer, J. Robertson, M. Wuttig, Resonant bonding in crystalline phase-change materials. *Nature Materials*. **7**, 653–658 (2008).

19. W. H. P. Pernice, H. Bhaskaran, Photonic nonvolatile memories using phase change materials. *Appl. Phys. Lett.* **101**, 171101 (2012).

20. C. Ríos, M. Stegmaier, P. Hosseini, D. Wang, T. Scherer, C. D. Wright, H. Bhaskaran, W. H. P. Pernice, Integrated all-photonic nonvolatile multi-level memory. *Nature Photon*. **9**, 725–732 (2015).

21. X. Li, N. Youngblood, C. Ríos, Z. Cheng, C. D. Wright, W. H. Pernice, H. Bhaskaran, Fast and reliable storage using a 5 bit, nonvolatile photonic memory cell. *Optica, OPTICA*. **6**, 1–6 (2019).

22. C. Rios, P. Hosseini, C. D. Wright, H. Bhaskaran, W. H. P. Pernice, On-Chip Photonic Memory Elements Employing Phase-Change Materials. *Advanced Materials*. **26**, 1372–1377 (2014).

23. Z. Cheng, C. Ríos, N. Youngblood, C. D. Wright, W. H. P. Pernice, H. Bhaskaran, Device-Level Photonic Memories and Logic Applications Using Phase-Change Materials. *Advanced Materials*. **30**, 1802435 (2018).

24. J. Feldmann, N. Youngblood, M. Karpov, H. Gehring, X. Li, M. Stappers, M. Le Gallo, X. Fu, A. Lukashchuk, A. S. Raja, J. Liu, C. D. Wright, A. Sebastian, T. J. Kippenberg, W. H. P. Pernice, H. Bhaskaran, Parallel convolutional processing using an integrated photonic tensor core. *Nature*. **589**, 52–58 (2021).

25. C. Ríos, N. Youngblood, Z. Cheng, M. Le Gallo, W. H. P. Pernice, C. D. Wright, A. Sebastian, H. Bhaskaran, In-memory computing on a photonic platform. *Science Advances*. **5**, eaau5759.

26. J. Feldmann, N. Youngblood, C. D. Wright, H. Bhaskaran, W. H. P. Pernice, All-optical spiking neurosynaptic networks with self-learning capabilities. *Nature*. **569**, 208–214 (2019).

27. C. Wu, H. Yu, S. Lee, R. Peng, I. Takeuchi, M. Li, Programmable phase-change metasurfaces on waveguides for multimode photonic convolutional neural network. *Nat Commun*. **12**, 96 (2021).

28. J. Zheng, Z. Fang, C. Wu, S. Zhu, P. Xu, J. K. Doylend, S. Deshmukh, E. Pop, S. Dunham, M. Li, A. Majumdar, Nonvolatile Electrically Reconfigurable Integrated Photonic Switch Enabled by a Silicon PIN Diode Heater. *Advanced Materials*. **32**, 2001218 (2020).

29. H. Zhang, L. Zhou, L. Lu, J. Xu, N. Wang, H. Hu, B. M. A. Rahman, Z. Zhou, J. Chen, Miniature Multilevel Optical Memristive Switch Using Phase Change Material. *ACS Photonics*. **6**, 2205–2212 (2019).

30. C. Ríos, Q. Du, Y. Zhang, C.-C. Popescu, M. Y. Shalaginov, P. Miller, C. Roberts, M. Kang, K. A. Richardson, T. Gu, S. A. Vitale, J. Hu, Ultra-compact nonvolatile photonics based on electrically reprogrammable transparent phase change materials. *arXiv:2105.06010 [cond-mat, physics:physics]* (2021) (available at http://arxiv.org/abs/2105.06010).

31. K. Kato, M. Kuwahara, H. Kawashima, T. Tsuruoka, H. Tsuda, Current-driven phase-change optical gate switch using indium–tin-oxide heater. *Appl. Phys. Express*. **10**, 072201 (2017).

32. H. Zhang, L. Zhou, J. Xu, N. Wang, H. Hu, L. Lu, B. M. A. Rahman, J. Chen, Nonvolatile waveguide transmission tuning with electrically-driven ultra-small GST phase-change material. *Science Bulletin*. **64**, 782–789 (2019).

33. H. Taghinejad, H. Taghinejad, S. Abdollahramezani, S. Abdollahramezani, A. A. Eftekhar, A. A. Eftekhar, T. Fan, A. H. Hosseinnia, O. Hemmatyar, A. E. Dorche, A. Gallmon, A. Adibi, ITO-based microheaters for reversible multi-stage switching of phase-change materials: towards miniaturized beyond-binary reconfigurable integrated photonics. *Opt. Express, OE*. **29**, 20449–20462 (2021).

34. J. Zheng, A. Khanolkar, P. Xu, S. Colburn, S. Deshmukh, J. Myers, J. Frantz, E. Pop, J. Hendrickson, J. Doylend, N. Boechler, A. Majumdar, GST-on-silicon hybrid nanophotonic integrated circuits: a nonvolatile quasi-continuously reprogrammable platform. *Opt. Mater. Express, OME*. **8**, 1551–1561 (2018).

35. C. Wu, H. Yu, H. Li, X. Zhang, I. Takeuchi, M. Li, Low-Loss Integrated Photonic Switch Using Subwavelength Patterned Phase Change Material. *ACS Photonics*. **6**, 87–92 (2019).

36. Q. Zhang, Y. Zhang, J. Li, R. Soref, T. Gu, J. Hu, Broadband nonvolatile photonic switching based on optical phase change materials: beyond the classical figure-of-merit. *Opt. Lett., OL*. **43**, 94–97 (2018).

37. J. Zheng, S. Zhu, P. Xu, S. Dunham, A. Majumdar, Modeling Electrical Switching of Nonvolatile Phase-Change Integrated Nanophotonic Structures with Graphene Heaters. *ACS Appl. Mater. Interfaces*. **12**, 21827–21836 (2020).

38. M. Delaney, I. Zeimpekis, H. Du, X. Yan, M. Banakar, D. J. Thomson, D. W. Hewak, O. L. Muskens, Nonvolatile programmable silicon photonics using an ultralow-loss Sb2Se3 phase change material. *Science Advances*. **7**, eabg3500.

39. M. Stegmaier, C. Ríos, H. Bhaskaran, C. D. Wright, W. H. P. Pernice, Nonvolatile All-Optical 1 × 2 Switch for Chipscale Photonic Networks. *Advanced Optical Materials*. **5**, 1600346 (2017).

40. M. Delaney, I. Zeimpekis, D. Lawson, D. W. Hewak, O. L. Muskens, A New Family of Ultralow Loss Reversible Phase-Change Materials for Photonic Integrated Circuits: Sb2S3 and Sb2Se3. *Advanced Functional Materials*. **30**, 2002447 (2020).

41. W. Dong, H. Liu, J. K. Behera, L. Lu, R. J. H. Ng, K. V. Sreekanth, X. Zhou, J. K. W. Yang, R. E. Simpson, Wide Bandgap Phase Change Material Tuned Visible Photonics. *Advanced Functional Materials*. **29**, 1806181 (2019).

42. Z. Fang, J. Zheng, A. Saxena, J. Whitehead, Y. Chen, A. Majumdar, Nonvolatile Reconfigurable Integrated Photonics Enabled by Broadband Low-Loss Phase Change Material. *Advanced Optical Materials*. **9**, 2002049 (2021).

43. Y. Zhang, J. B. Chou, J. Li, H. Li, Q. Du, A. Yadav, S. Zhou, M. Y. Shalaginov, Z. Fang, H. Zhong, C. Roberts, P. Robinson, B. Bohlin, C. Ríos, H. Lin, M. Kang, T. Gu, J. Warner, V. Liberman, K. Richardson, J. Hu, Broadband transparent optical phase change materials for high-performance nonvolatile photonics. *Nat Commun*. **10**, 4279 (2019).

44. S. Kim, G. W. Burr, W. Kim, S.-W. Nam, Phase-change memory cycling endurance. *MRS Bulletin*. **44**, 710–714 (2019).

45. M. R. Watts, J. Sun, C. DeRose, D. C. Trotter, R. W. Young, G. N. Nielson, Adiabatic thermo-optic Mach–Zehnder switch. *Opt. Lett.* **38**, 733 (2013).

46. C. Kieninger, C. Kieninger, C. Kieninger, Y. Kutuvantavida, Y. Kutuvantavida, D. L. Elder, S. Wolf, H. Zwickel, M. Blaicher, J. N. Kemal, M. Lauermann, S. Randel, W. Freude, L. R. Dalton, C. Koos, C. Koos, Ultra-high electro-optic activity demonstrated in a silicon-organic hybrid modulator. *Optica, OPTICA*. **5**, 739–748 (2018).

47. C. Qiu, W. Gao, R. Soref, J. T. Robinson, Q. Xu, Reconfigurable electro-optical directed-logic circuit using carrier-depletion micro-ring resonators. *Opt. Lett., OL*. **39**, 6767–6770 (2014).

48. C. Haffner, A. Joerg, M. Doderer, F. Mayor, D. Chelladurai, Y. Fedoryshyn, C. I. Roman, M. Mazur, M. Burla, H. J. Lezec, V. A. Aksyuk, J. Leuthold, Nano–opto-electro-mechanical switches operated at CMOS-level voltages. *Science*. **366**, 860–864 (2019).

49. A. Yariv, Coupled-mode theory for guided-wave optics. *IEEE Journal of Quantum Electronics*. **9**, 919–933 (1973).

50. L. Zhuang, C. G. H. Roeloffzen, M. Hoekman, K.-J. Boller, A. J. Lowery, Programmable photonic signal processor chip for radiofrequency applications. *Optica, OPTICA*. **2**, 854–859 (2015).

51. Z. Cheng, C. Ríos, W. H. P. Pernice, C. D. Wright, H. Bhaskaran, On-chip photonic synapse. *Science Advances*. **3**, e1700160.

# Acknowledgments

**Funding:** The research is funded by National Science Foundation (NSF-1640986, NSF-2003509), ONR-YIP Award, DRAPER Labs and Intel. Part of this work was conducted at the Washington Nanofabrication Facility / Molecular Analysis Facility, a National Nanotechnology Coordinated Infrastructure (NNCI) site at the University of Washington with partial support from the National Science Foundation via awards NNCI-1542101 and NNCI-2025489. **Author contributions:** J.Z. and A.M. conceived the project. R.C. simulated the nonvolatile switches, fabricated the samples, performed optical characterizations and data analysis. J.Z. optimized and simulated all the devices, planned the layout, and helped with data analysis. Z.F. fabricated and measured test structures, and helped with other fabrications, optical measurements, and data analysis. J.F. helped with device characterization. P.X. helped with the FDTD simulation and discussed with J.Z. about the design. A.M. supervised and planned the project. R.C. wrote the manuscript with input from all the authors. **Competing interests:** Authors declare that they have no competing interests. **Data and materials availability:** The data that support the findings of this study are available from the corresponding author upon reasonable request.

# Broadband nonvolatile electrically programmable silicon photonic switch Supplementary Materials

Rui Chen,[1] Zhuoran Fang,[1] Johannes E. Fröch,[1] Peipeng Xu,[2] Jiajiu Zheng, [1]* Arka Majumdar[1,3]*

[1]Department of Electrical and Computer Engineering, University of Washington, Seattle, WA 98195, USA

[2]Faculty of Electrical Engineering and Computer Science, Key Laboratory of Photoelectric Materials and Devices of Zhejiang Province, Ningbo University, Ningbo, 315211, China

[3]Department of Physics, University of Washington, Seattle, WA 98195, USA

* jjzno1@gmail.com, arka@uw.edu

## The PDF includes

# Section 1. Measurement setup

The experimental setup described in the *Method* section is shown graphically in *Fig. S1*.

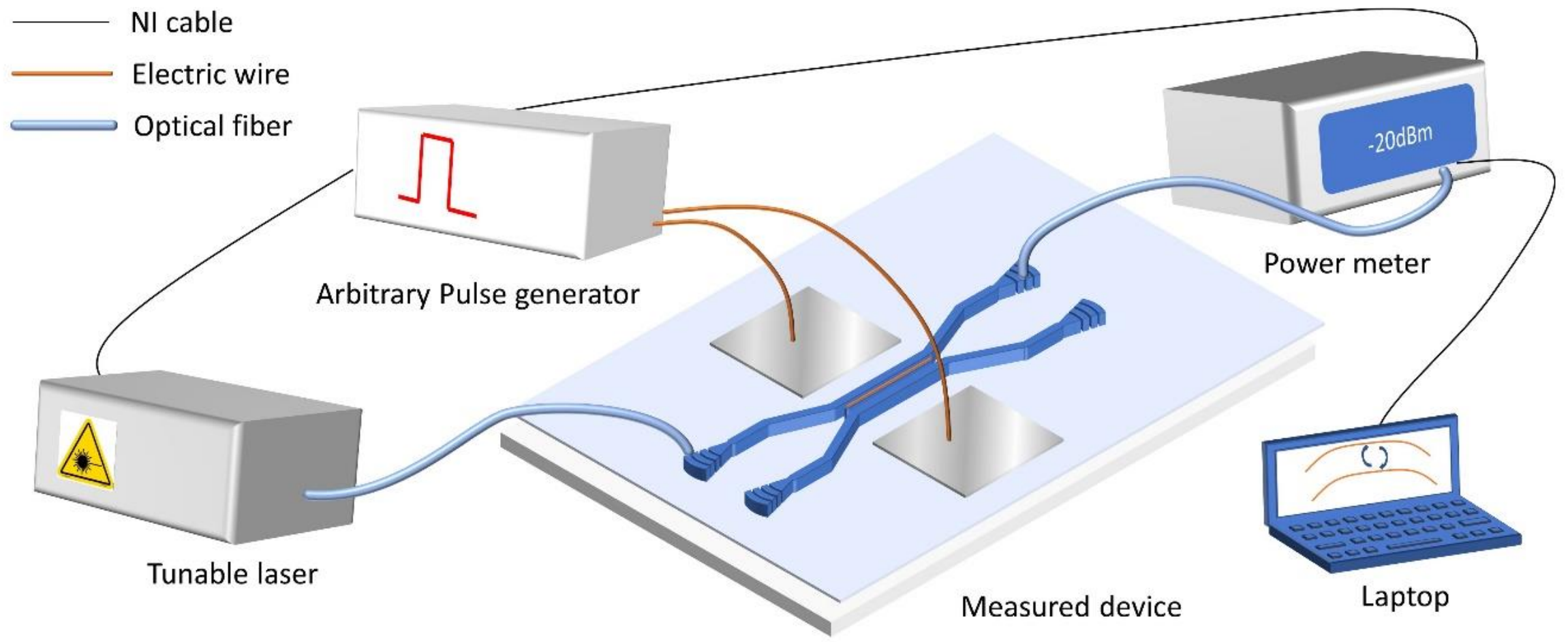

Figure S1. **Measurement setup schematics**.

# Section 2. Optical simulation results

In this section, we present a detailed design procedure of our 1 × 2 and 2 × 2 electrically controllable nonvolatile broadband switches.

## Section 2.1 GST optical characterization

GST has a large loss in the crystalline state at 1550 nm, as shown in *Fig. S2*, which is measured and fitted with ellipsometry. Note that the extinction coefficient $\kappa$ is proportional the optical loss $\alpha(dB/m)$ as

$$\alpha = 10\,\lg(e) \cdot \frac{4\pi\kappa}{\lambda}$$

( 1 )

We simulate the fundamental TE mode of a hybrid GST silicon rib waveguide (HW) using Lumerical MODE. The results are shown in *Fig. S4(d)*, where a high loss of 5.5 dB/μm when the GST is in the crystalline state is estimated. Directly using GST in photonic switches will introduce huge insertion loss. For example, if GST is used in an MZI, then to achieve a π-phase shift, a length of $L = \frac{\lambda_0}{2\Delta n} = 4.2\mu m$ would be required. This will then give a very high loss of 23.1 dB compared to <1 dB loss using the three waveguide structure in the following sections.

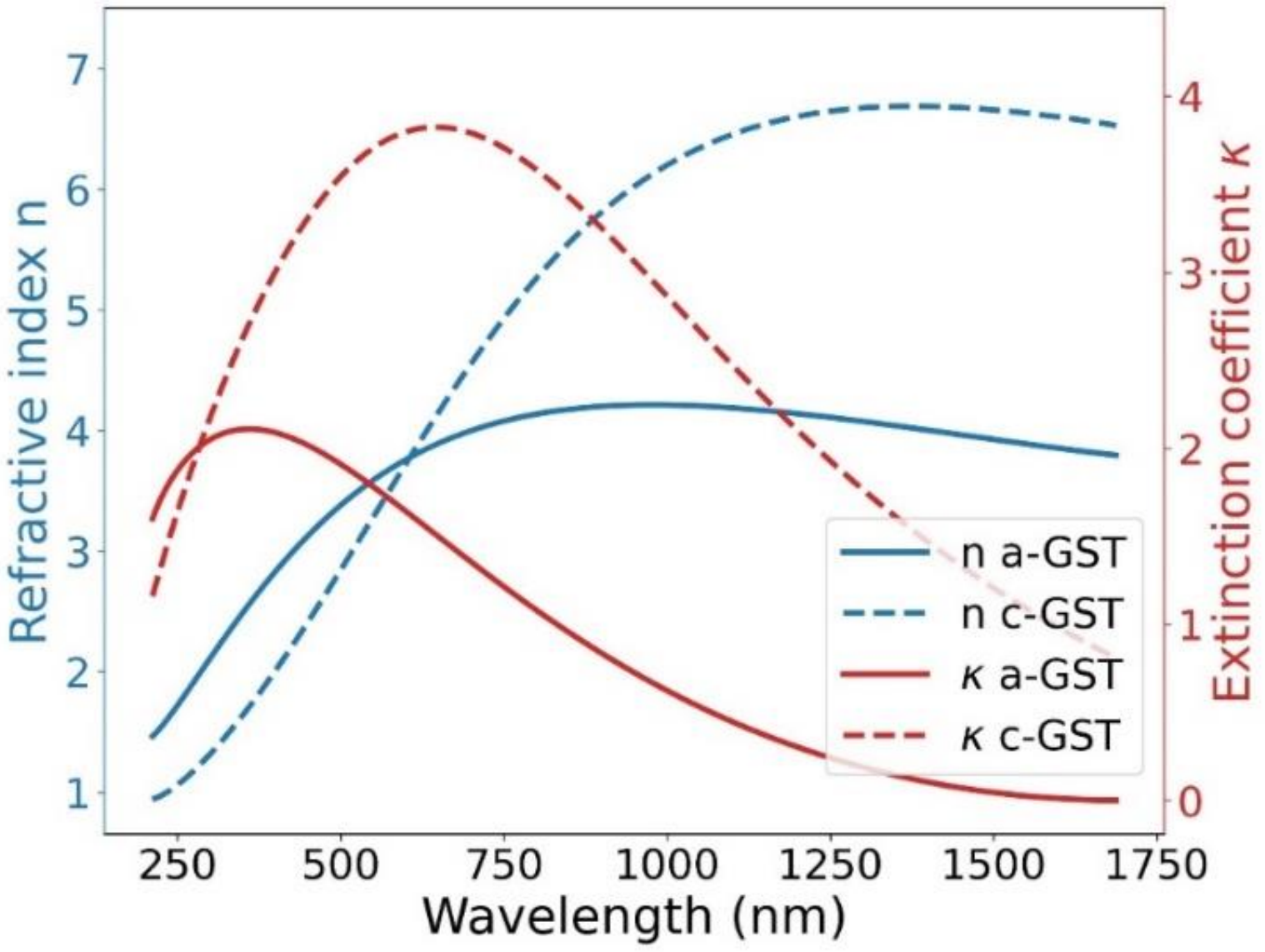

Figure S2. **The complex refractive index of a/c-GST measured by ellipsometry.**

## Section 2.2 1 × 2 switch design

We design the 1 × 2 switch so that the light can only couple from the input bare silicon waveguide (SW) to the hybrid (HW) when the GST is in the amorphous state, where only a small loss is introduced, as shown in *Fig. 1* and *S3*.

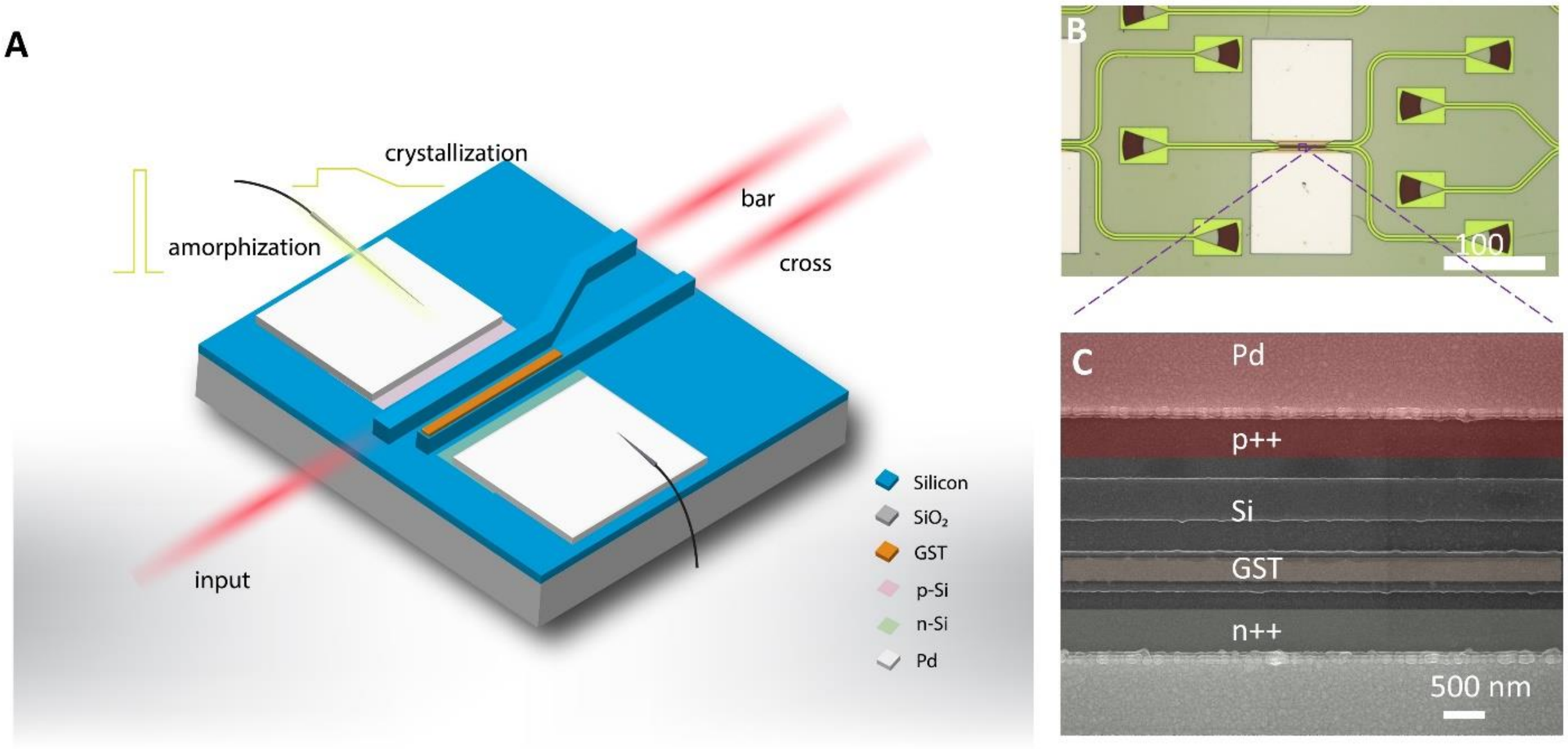

Figure S3. **Nonvolatile electrically controllable 1 × 2 broadband photonic switch schematic and images** (a) device schematic (b) The microscope and (c) SEM images of the device. In the SEM image, the GST thin film, p++, and n++ regions are indicated by orange, blue, and green fake colors.

The design procedure of the switches is similar to (*13*) as follows. The HW width was optimized so that when GST is in the amorphous state, the HW and SW can couple to each other most strongly, which happens when the two share the same effective index. As shown in *Fig. S4(a)*, a 408 nm a-GST HW has the same effective index with a 450 nm SW. Due to the significant refractive index change when the GST is switched into a crystalline state, a large phase mismatch would be induced. Thus the coupling would be weakened.

Then the supermodes of the HW and SW with a gap $g$ are simulated. The coupling length $L_c$ and the loss when GST is in the amorphous/crystalline state are determined when sweeping $g$, where the coupling length is determined with coupled mode theory(*49*), using the following formula:

$$L_c^a = \frac{\lambda_0}{2(n_{even} - n_{odd})}$$

( 2 )

Where $\lambda_0$ is the free-space wavelength, $n_{even}$ and $n_{odd}$ are effective index for even and odd mode and $L_c^a$ is the coupling length with a-GST. We note that the loss is determined by decomposing the input field into the supermodes and calculating the loss separately, as below:

$$T^{a,c} = \sum_j T_j^{a,c} = \sum_j \left( A_j^{a,c} \exp\left( \frac{i2\pi\kappa_j^{a,c}}{\lambda} L_c^a \right) \right)^2$$

( 3 )

$$loss^{a,c} = 10 \log T^{a,c}$$

( 4 )

Where $T^{a,c}$ is the transmission and the superscript indicates the same formula applies to both amorphous and crystalline GST, $T_j^{a,c}$ is the transmission of the $j^{th}$ supermode, $A_j^{a,c}$ is the expansion coefficient of the input field into the $j^{th}$ supermode field, $\kappa_j^{a,c}$ is the imaginary part of the effective index of the $j^{th}$ supermode. The crosstalk of the device can be calculated with similar formula as well. This formula was implemented in the Lumerical MODE solver.

The result is shown in *Fig. S4(c)*. As the *g* increases, the coupling length increases, the loss in c-state decreases because of a weaker mode interaction. However, since the amorphous state GST still has a finite amount of loss, a longer coupling length leads to a larger insertion loss. We selected *g* of 450nm to get an insertion loss smaller than 0.5 dB and a relatively compact device footprint of $L_c$ = 42μm. The supermode simulation results are shown in *Fig. S4(e)*. With a-GST, the supermode extends in both waveguides, indicating a strong coupling. While with c-GST, the supermodes are highly localized in each of the waveguides, indicating weak coupling. Thus, the light bypasses the high absorption loss of c-GST.

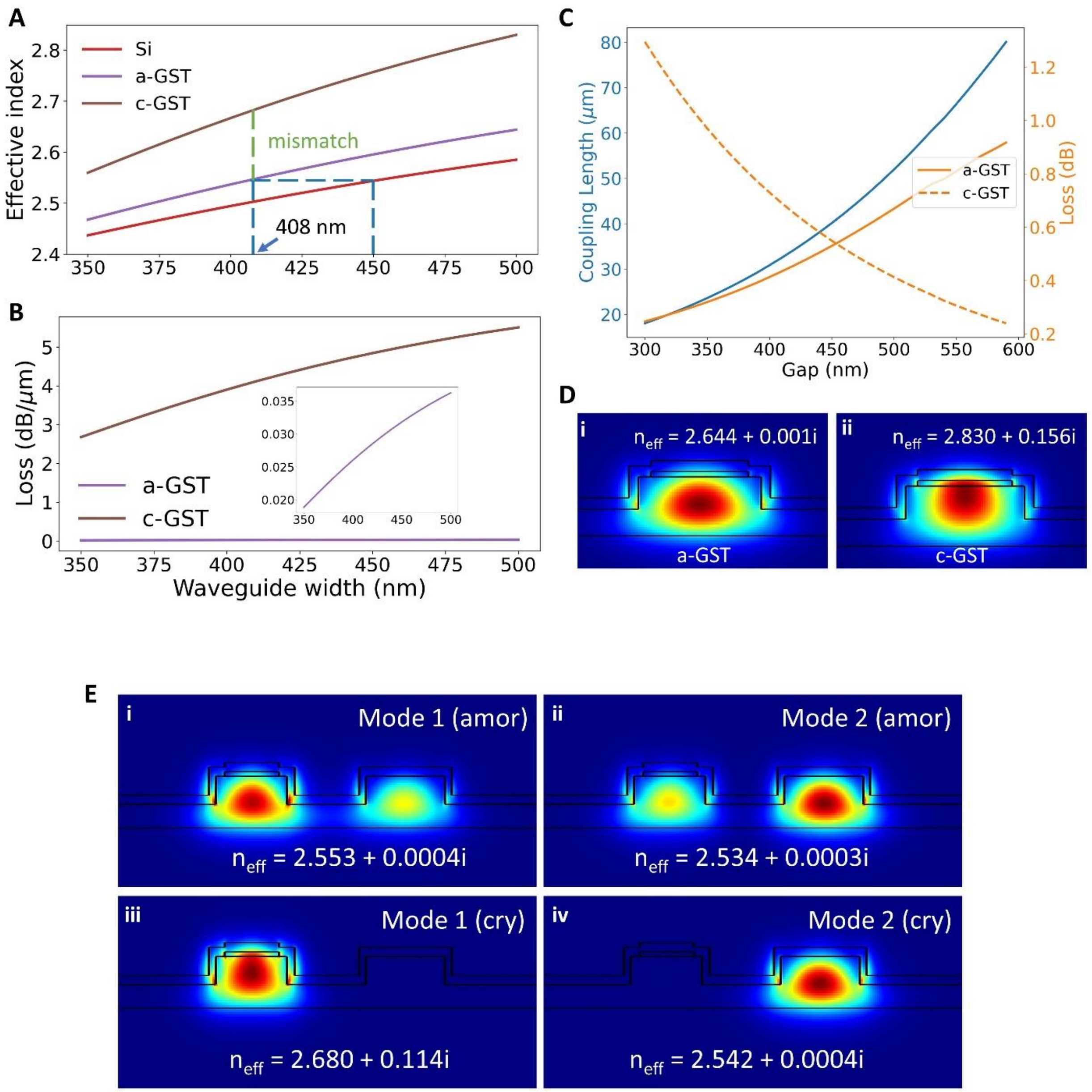

Figure S4. **Mode simulation results for the 1 × 2 switch.** (a) effective indices of a bare silicon waveguide, a-GST HW, and c-GST HW changing with the waveguide width. (b)Effective loss of a/c-GST HW, the inset shows the zoomed-in loss of a-GST HW. (c) Mode simulation result of 1 × 2 switch coupling length and insertion loss with a/c-GST changing with gap. (d) Mode simulation results for (i)a-GST and (ii)c-GST HW. (e) Mode simulation results for 1 × 2 switch supermodes. (i)(ii) First and second order mode when GST is in amorphous state. (iii)(iv) First and second order mode when GST is in crystalline state.

## Section 2.3 2 × 2 switch design

The first three TE supermodes are considered for the 2 × 2 switch design. To achieve the best coupling between the three waveguides, a phase-matching condition needs to be satisfied:

$$2n_2 = n_1 + n_3 \tag{5}$$

By fixing the width of SWs to 450 nm and sweeping the width of the middle waveguide, an optimal HW width of 409 nm is determined, as shown in *Fig S5(a)*. The minor difference from the previously determined HW width for 1 × 2 switch can be attributed to numerical error.

The coupling length and the insertion loss are calculated with different gap *g*, as shown in *Fig S5(b)*. An optimal gap of 450 nm is selected for a short coupling length, as well as a small insertion loss of 0.33/0.65dB for a/c-GST, respectively. The coupling length of a three waveguide coupler is calculated by the following formula when the phase-matching condition is satisfied:

$$L_c^a = \frac{\lambda}{2(n_1 - n_2)} \tag{6}$$

Finally, we show the optical modes in *Fig. S5(c)*. One can see a strong coupling between the HW and SW when GST is in the amorphous state, as in *Fig. S5(c)(i, iii)*. However, when the GST is switched into a crystalline state, the modes are localized in SW or HW, as in *Fig. S5(c)(iv, v, vi)*, indicating a very weak coupling. Although the effective index of mode 1 (cry) has a large imaginary part, since light can't get coupled into mode 1 if excited from SW, this high loss mode is bypassed.

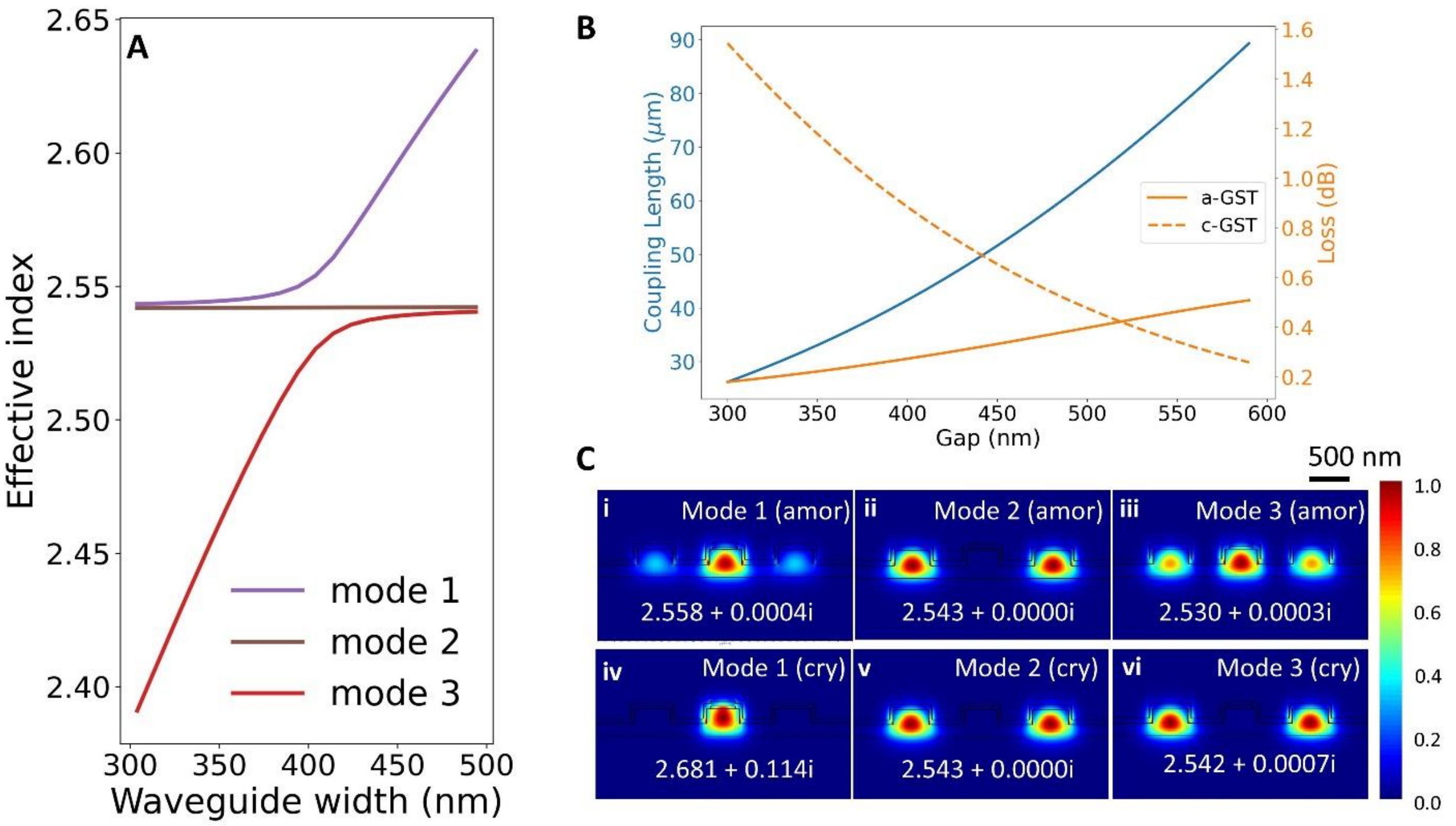

Fig S5. **Mode simulation results for the 2 × 2 switch.** (a) supermode effective indices changing with the HW width. (b) coupling length and loss with a/c-GST changing with the gap. (c) Mode simulation results. (i)(ii)(iii) First, second, and third-order modes when GST is in the amorphous state, (iv)(v)(vi) First, second, and third-order modes when GST is in the crystalline state.

The designs of the 1 × 2 and 2 × 2 switch are verified with Lumerical FDTD as in *Fig S6*. A larger insertion loss of around 1.0 dB compared to the 0.5 dB loss in the MODE solver can be attributed to extra loss introduced by the S-bends and can be further reduced by a better S-bend design.

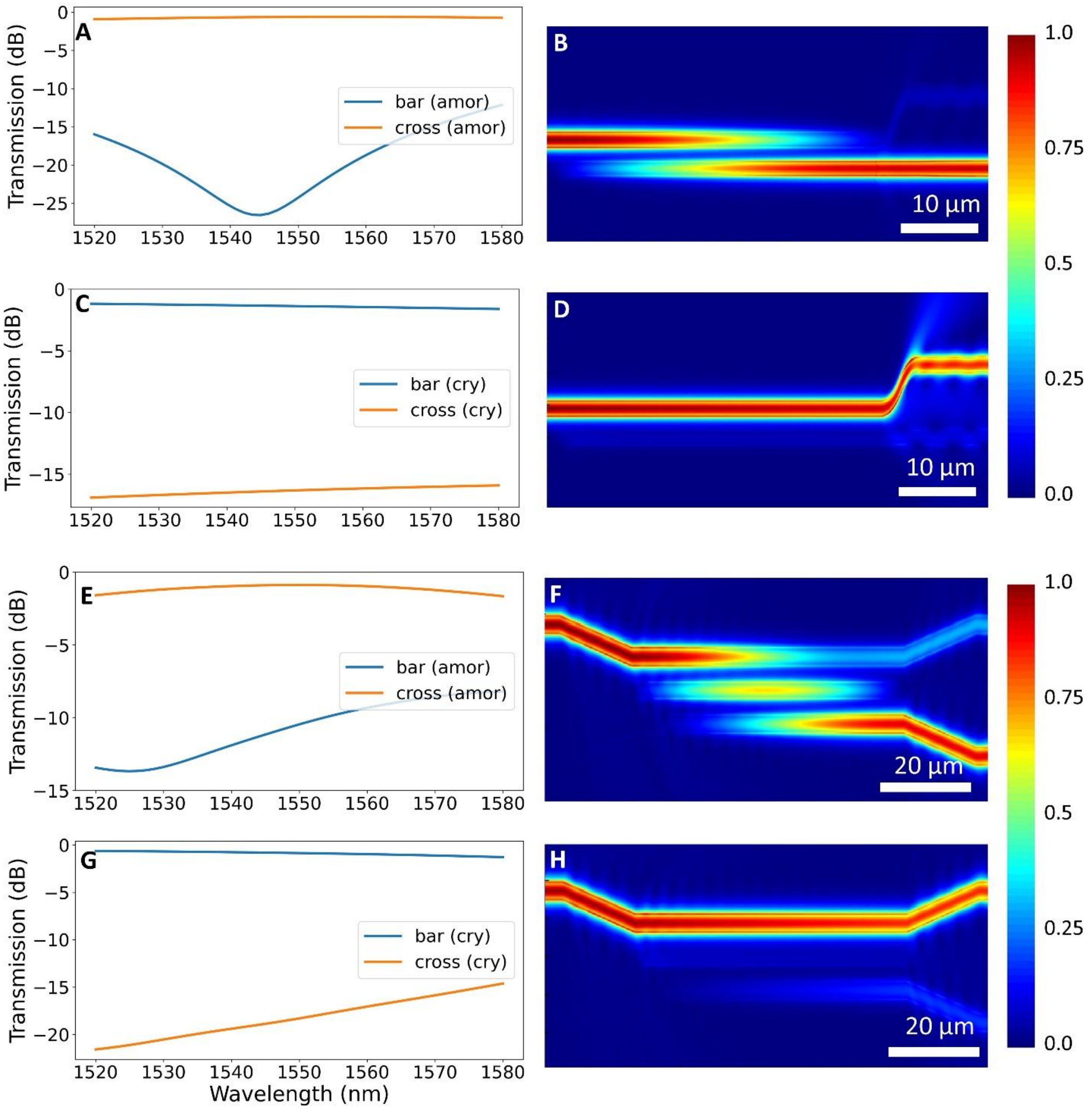

Figure S6. **FDTD simulation results for both 1 × 2 and 2 × 2 switches.** Transmission spectrum and field distribution of (a)(b) a-GST 1 × 2 switch, (c)(d) c-GST 1 × 2 switch, (e)(f) a-GST 2 × 2 switch, (g)(h) c-GST 2 × 2 switch.

## Section 2.4 Fabrication variation tolerance analysis

In this section, we simulate how the insertion loss and extinction ratio in both amorphous and crystalline state changes with respect to various device parameters, such as the waveguide width, the GST width, the GST misalignment, the etch depth and the gap. The results are summarized in *Fig S7*. The fabrication variation in waveguide width, etching depth and GST thickness can lead to significant higher insertion loss and lower extinction ratio.

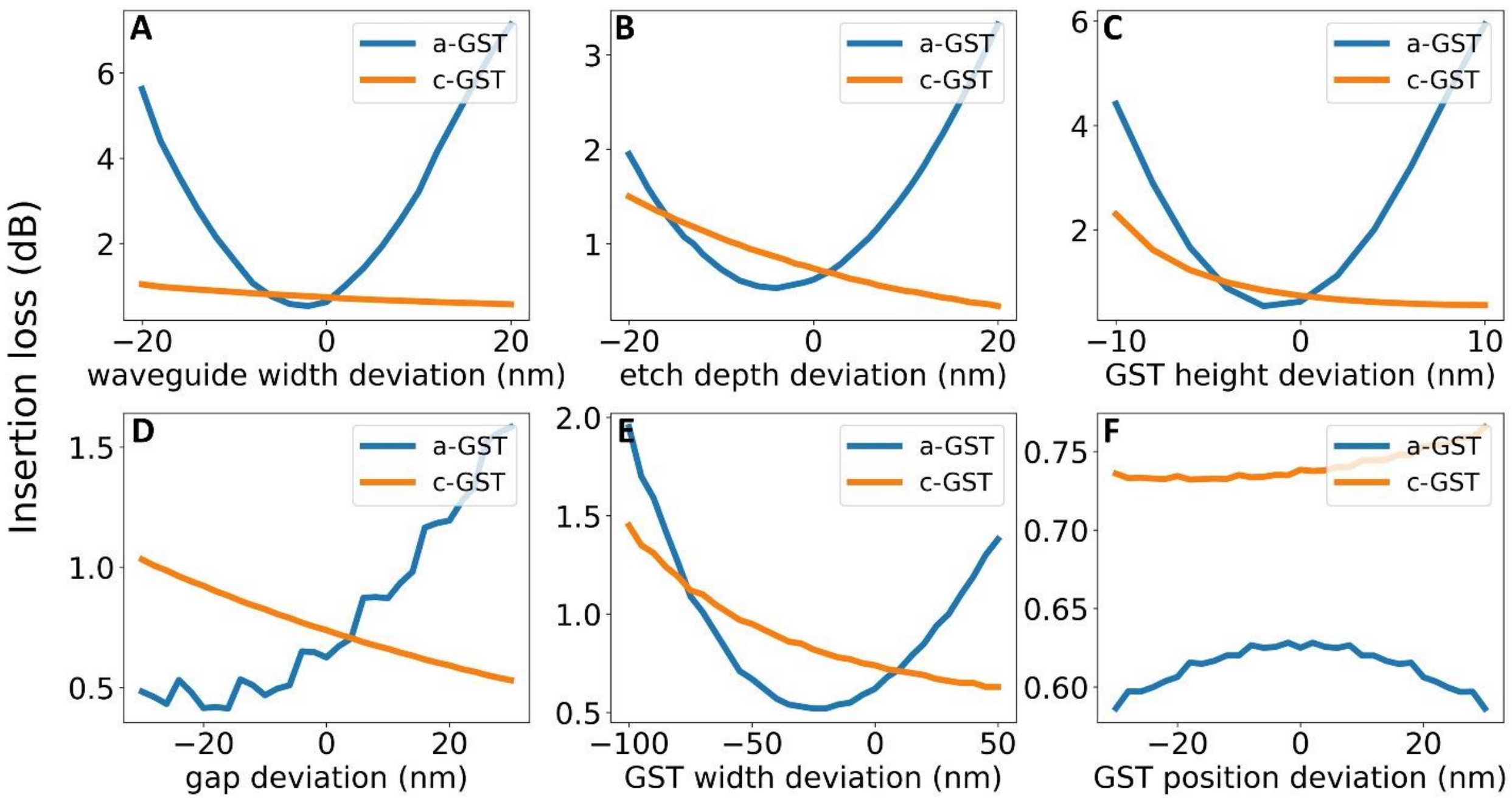

Figure S7. **Insertion loss with respect to fabrication variation study on the 2 × 2 switch.** (a) Waveguide width deviation (b) etch depth deviation (c) GST height deviation (d) gap deviation (e) GST width deviation (f) GST position deviation from the nominal values.

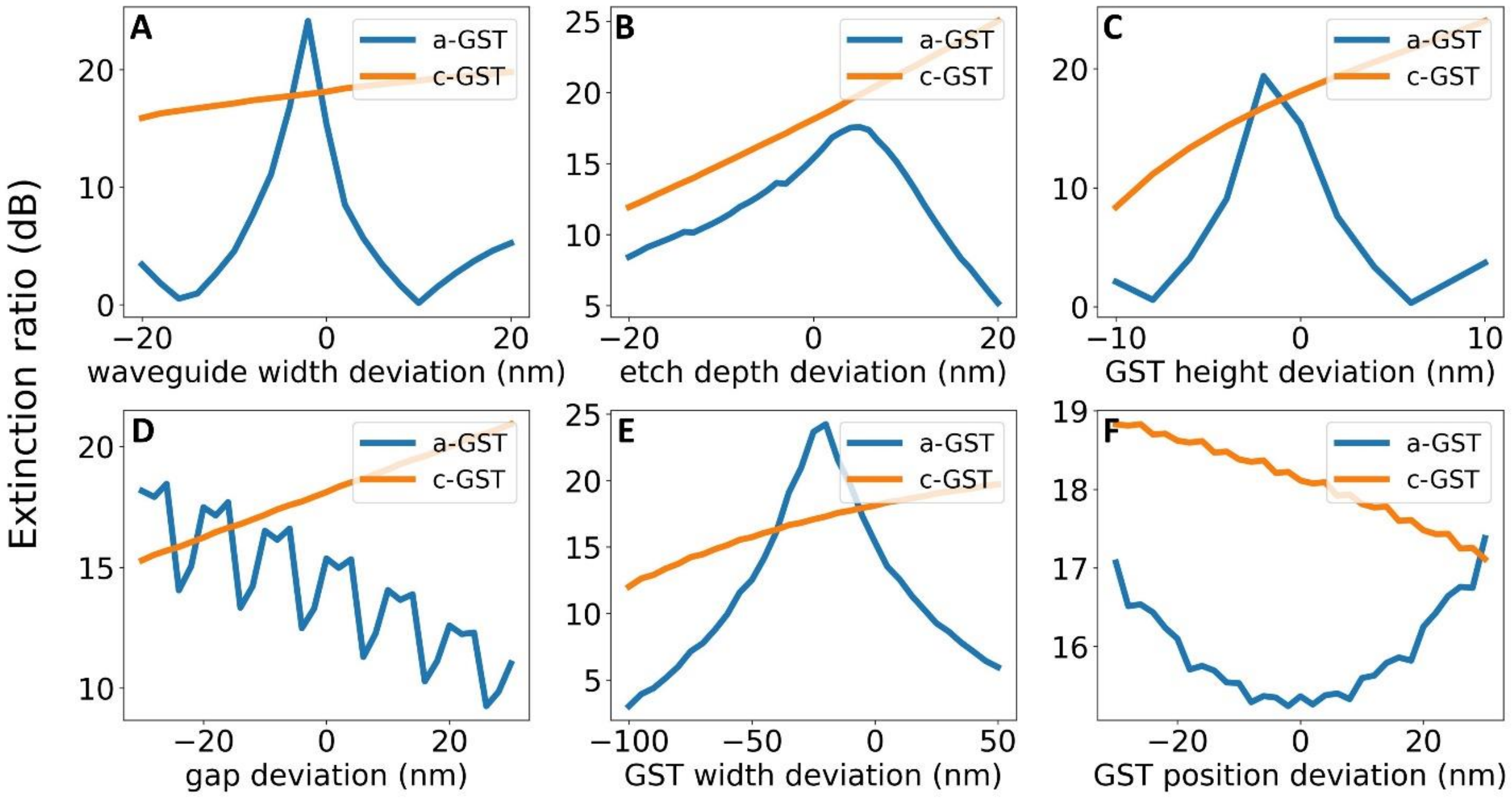

Figure S8. **Extinction ratio with respect to fabrication variation study on the 2 × 2 switch.** (a) Waveguide width deviation (b) etch depth deviation (c) GST height deviation (d) gap deviation (e) GST width deviation (f) GST position deviation from the nominal values.

# Section 3. Heat transfer simulation

The time-dependent heat transfer performance was studied with a cross-coupled electro-thermal model in COMSOL Multiphysics. The detailed description of the model can be found in(*28*, *37*).

For amorphization, a short sharp pulse with high peak voltage is favorable for the melting and quenching process, where the temperature of GST increases above its melting point ($T_m$ ~ 616°C) and decreases rapidly below the glass transition temperature ($T_g$ ~ 155°C) to form the amorphous phase. The amorphous GST can be crystallized through nucleation and growth process, by heating it above the $T_g$ but below $T_m$, and gradually cooling it down. The amorphization condition was identified as pulse duration of 100 ns, and peak amplitude of 6.8(8.4) V for the 1 × 2 (2 × 2) switches. To crystallize the GST, a pulse with a duration of 50 µs and peak amplitude of 2.4 (2.8) V. The larger peak amplitude required for the 2 × 2 switch is due to the wider intrinsic region, thus a larger resistance and heating volume. *Fig. S9(a) and (d)* shows the time dependent temperature result when the amorphization or crystallization pulses are applied. The experimental pulses have higher voltages because of nonideal contact and load impedance mismatch.

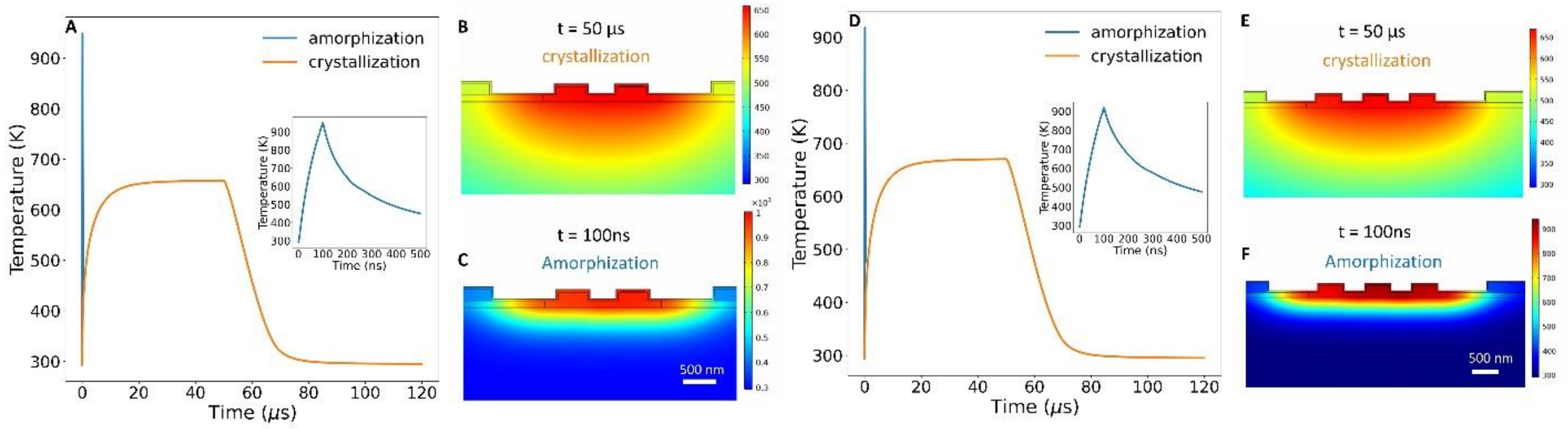

Figure S9. **Transient heat transfer simulation results.** (a)(d) The time-dependent temperature of the GST. The blue and orange lines are amorphization and crystallization pulses, respectively. The insets show zoom-in figures of the amorphization pulse. (b)(e) Temperature distribution at 50 µs for crystallization. (c)(f) Temperature distribution at 100 ns for amorphization. (a-c) 1 × 2 switch. (d-f) 2 × 2 switch.

# Section 4. Additional measurement results

In this section, we present additional measurements results on the 2 × 2 switch. We also present the measurement results of 1 × 2 and 1 × 1 switch.

## Section 4.1 more measurements on 2 × 2 switch

We show the full cyclability test results (~9000 switching events) in *Fig. S10*. The transmission contrast dropped significantly between 6000th and 6750th switching events. Still, when we increase the voltage amplitude from 14.0 V to 14.3V after the 6750th event, the transmission contrast gradually increases, as shown in *Fig. S10(d)*. This is interesting as the GST 'healed' itself gradually and provided a larger contrast.

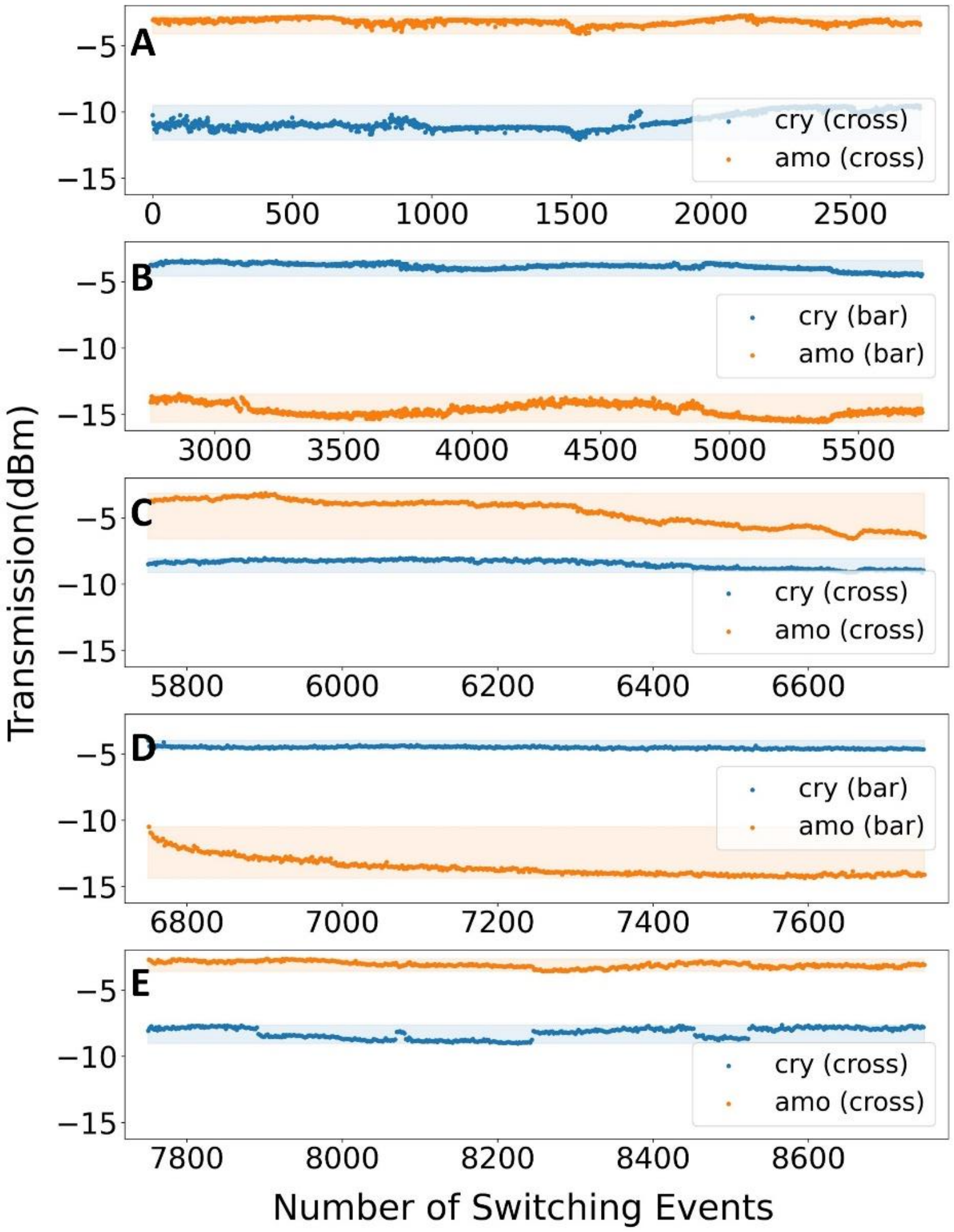

Figure S10. **Full endurance test results for 9000 switching events.**

We performed a continuous time-dependent transmission measurement on the device. The pulse generator sent set and reset pulses alternatively with an interval of 2 seconds when the power meter continuously measured the optical transmission from the cross/bar port as in *Fig. S11(a, b)*. After each pulse, the transmission changed drastically, then retained during the pulse intervals, indicating the change is nonvolatile. The minor transmission fluctuation between two pulses can be attributed to slight coupling condition variation when measuring. This experiment demonstrates the reversible and nonvolatile tuning capability of the 2 × 2 switch. The device current-voltage (I-V) characteristics were measured before and after the cycling test. The I-V curves, as in *Fig. S11(c)*, show a minor change during the cyclability test. *Fig. S11(d, e)* show the transmission spectra during the cyclability test, when the GST is in amorphous and crystalline states, respectively. The switch with a-GST has less performance degradation than that with c-GST. This can be attributed to material degradation and thermal reflow of the GST during repeated phase transition, as shown in *Fig. S16*.

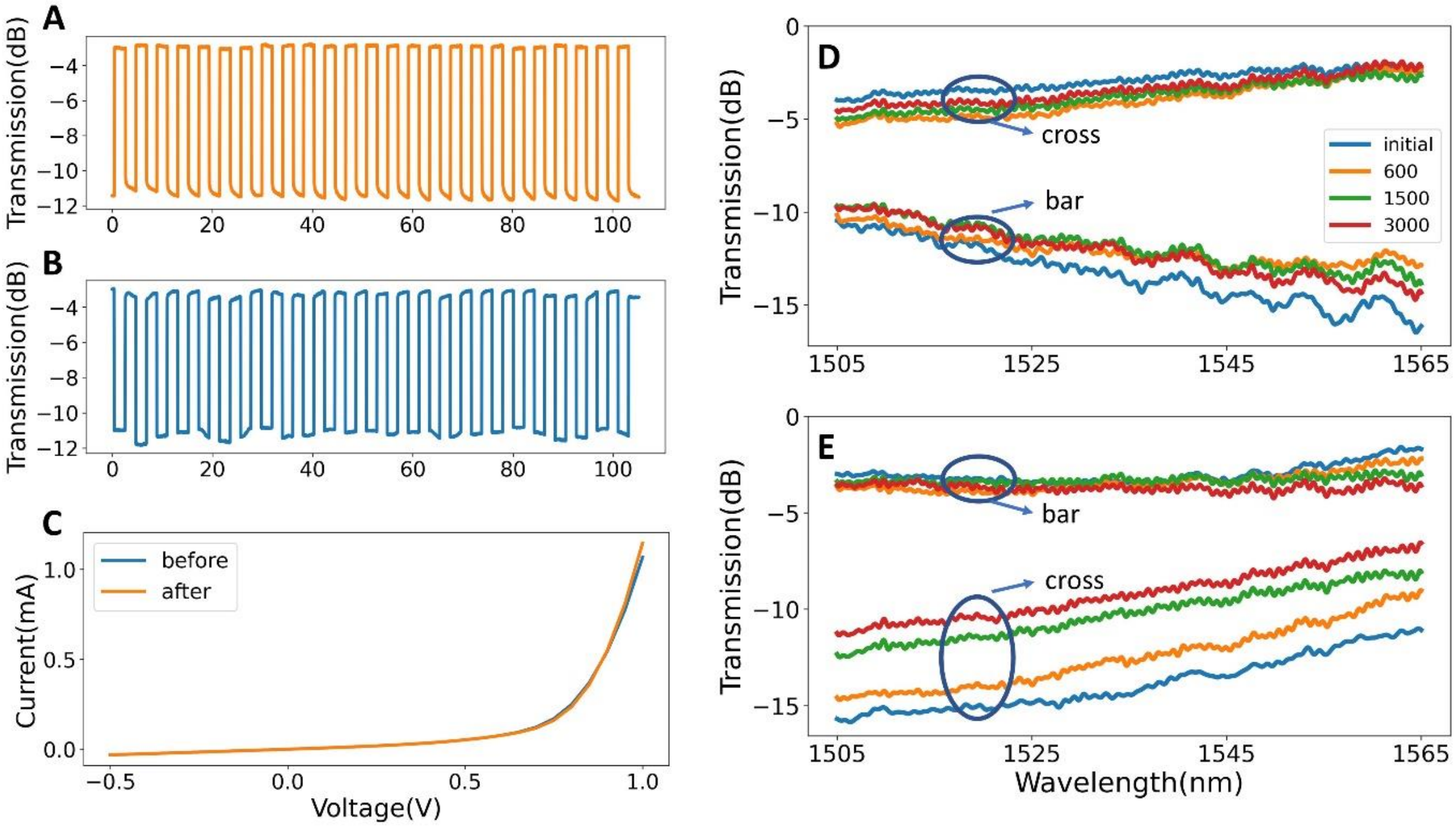

Figure S11. **Additional measurement results for the 2 × 2 switch.** Continuous time-dependent transmission measurement for (a) bar and (b) cross ports. (c) I-V curve before and after the cyclability test. (d)(e) The transmission spectrum after 0, 600, 1500, and 3000 cycles for (d) amorphous and (e) crystalline GST.

## Section 4.2 Measurement results for 1 × 2 switch

The 1 × 2 switch exhibited similar performance as the 2 × 2 switch. It has an extinction ratio of 9 dB, an insertion loss of 4 dB, nonvolatility, and large endurance of 8000 switching events, as in *Fig. S12* and *S13*. The high loss in this design can also be attributed to fabrication imperfection. We also note that 1 × 2 design inherently has a higher overall loss than 2 × 2 design, as shown in Fig. S4(C) and Fig. S5(B).

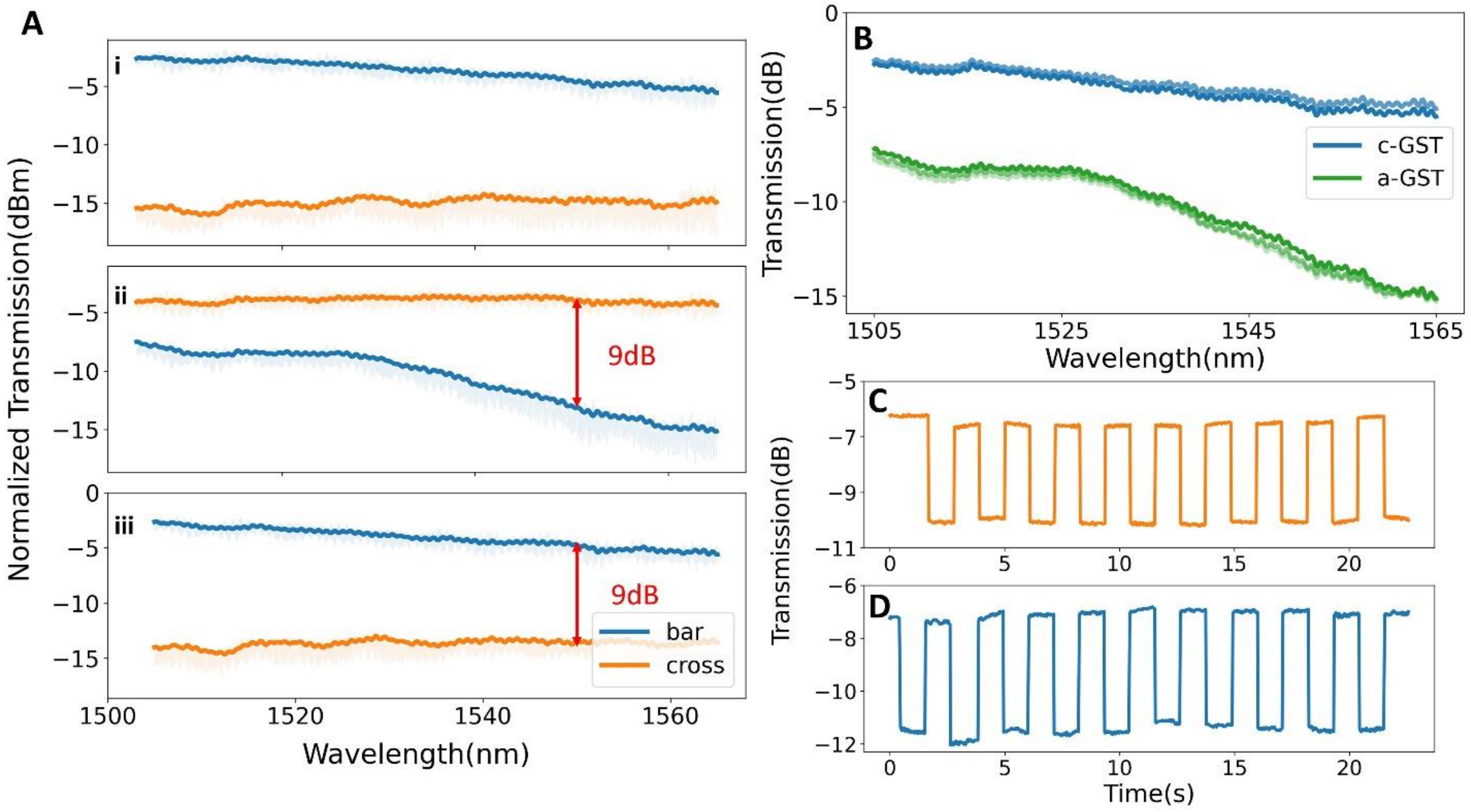

Figure S12. **Optical transmission spectra measurement results for the 1 × 2 switch.** (a) The transmission spectrum (i) initially when GST is in the crystalline state, (ii) after sending an amorphization pulse, (iii) after sending a crystallization pulse. (b) Transmission spectrum of 3 cycles amorphization and crystallization at the bar port. The spectra overlap well, showing a reliable and reversible switching performance. (c)(d) The continuous-time measurement result for (c) cross and (d) bar ports.

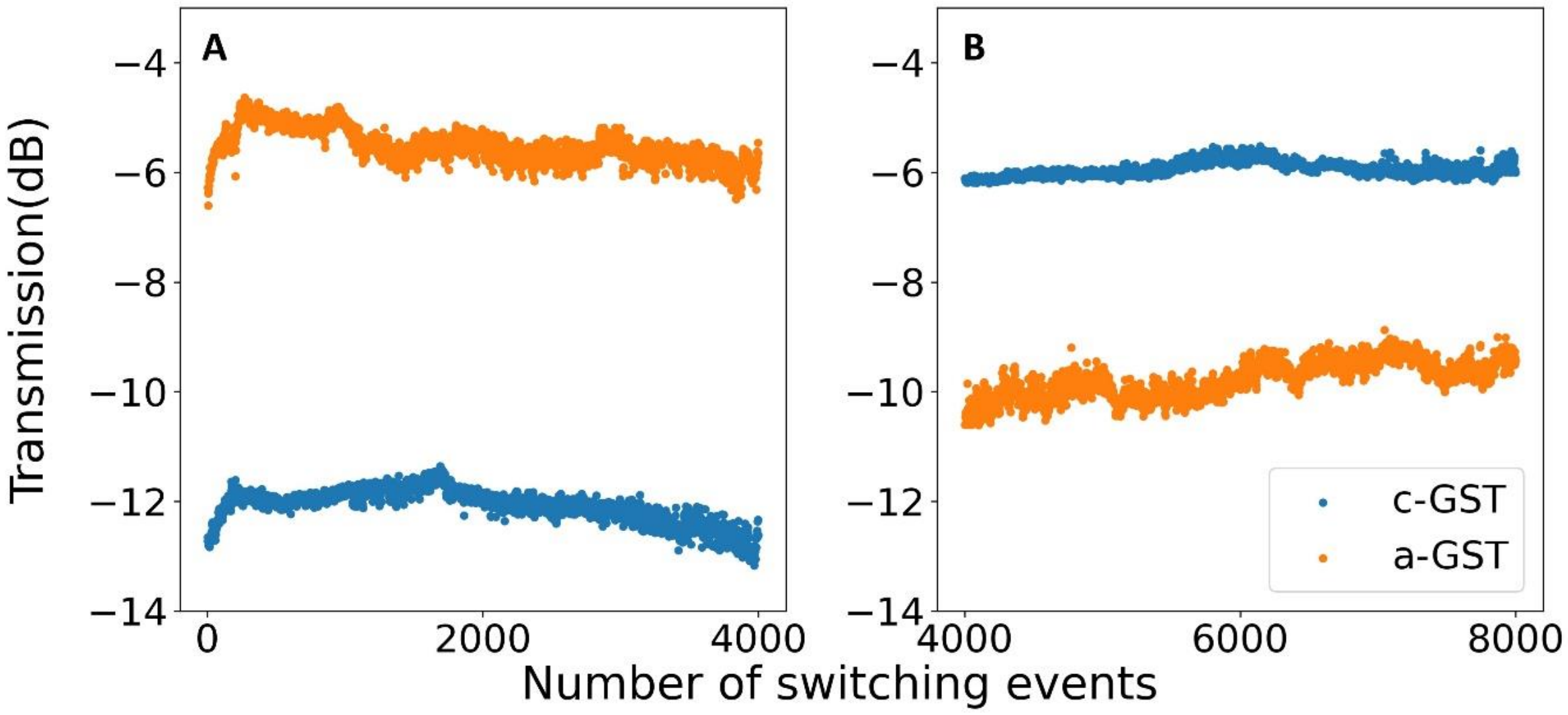

Figure S13. **The endurance test results for the 1 × 2 switch** at (a) cross and (b) bar port. 8000 switching events are shown without significant performance degradation at 1550nm. The degradation in extinction ratio and insertion loss compared to Fig. S12 is mainly due to material ablation when testing the switching pulse conditions.

## Section 4.3 Measurement results for waveguide switch

We also fabricated 10-µm-long GST on silicon waveguide switches, as shown in *Fig. 14(a)* and *(b)*. By switching the GST between amorphous/crystalline (low-loss/high-loss) states, the optical switch is controlled on/off, with an extinction ratio larger than 15 dB. We show that this device can be switched with ten distinct transmission levels. This was accomplished by first complete crystallizing (2.2 V peak voltage, 30 µs duration and 10 µs falling edge) the GST and then applying gradually increased voltage (peak volage starts from 6.5V, with an increment step of 0.1V, and 10 total steps, 200 ns pulse duration and 8 ns falling edge) for partial amorphization, as in *Fig. S14(c)*. We repeated the multilevel switching test three times at a wavelength of 1550 nm, and each time a multilevel performance can be observed, shown in *Fig. S14(e)*. We further demonstrate more than 2000 switching events via sending in amorphization (2.2V) and crystallization (7.4V) pulses alternately, and the results are shown in *Fig. S14(d)*. The waveguide switch exhibits a large extinction ratio between the ON and OFF state after more than 2000 switching events, showing excellent endurance. We note that the transmission level becomes stable after around 600 switching events due to the larger crystalline domain forming from the initial 'conditioning' steps(*20*).

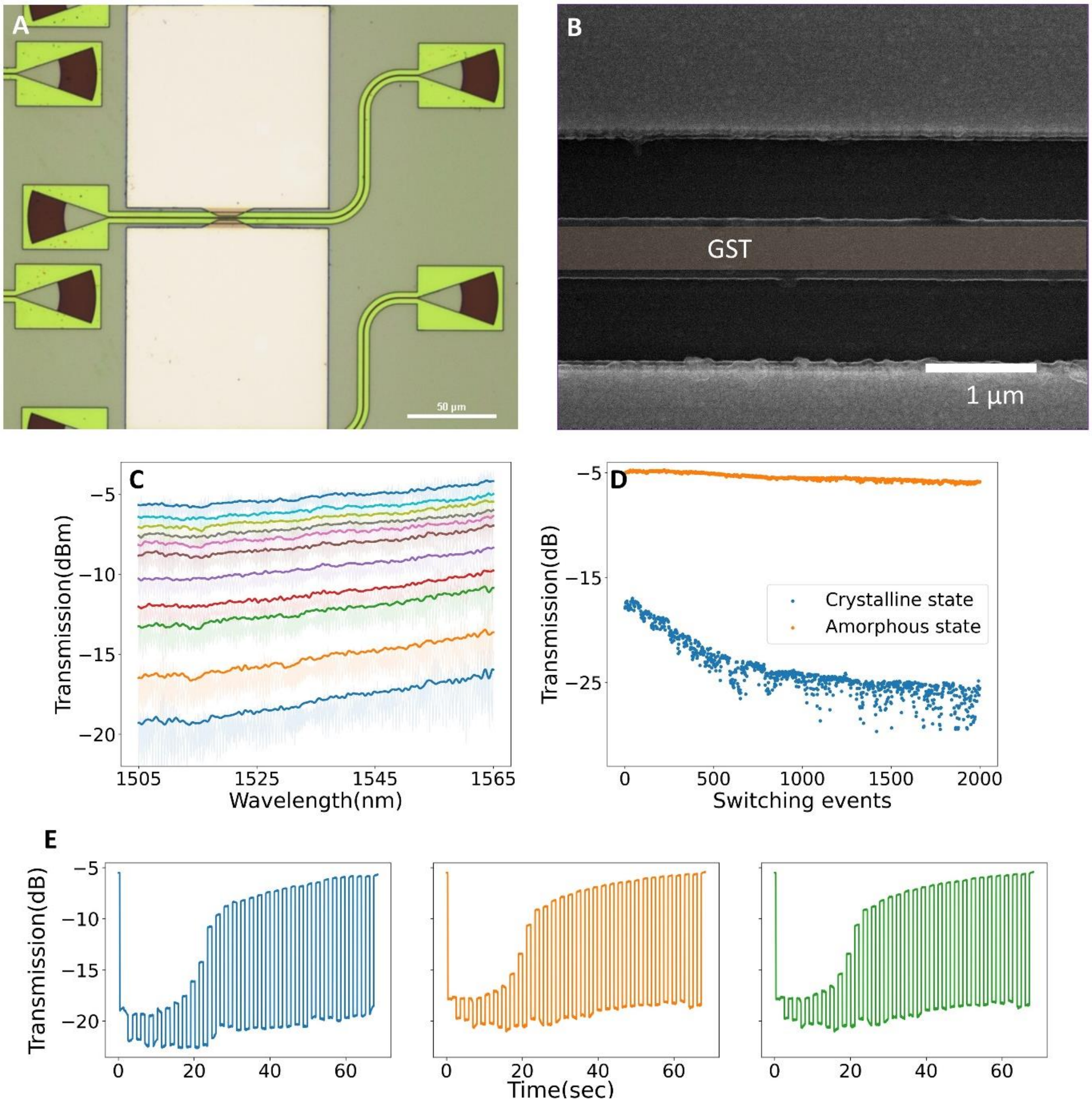

Fig S14. **Measurement results for the waveguide switch.** (a) Microscope and (b) SEM image of the waveguide, where the GST thin film is indicated by the orange fake color. (c) Multilevel switching results. (d) cyclability test results. (e) Three rounds of multilevel switching results.

# Section 5. Thermal stability test results

One advantage of a broadband device is its thermal stability. Here we measured the transmission spectrum of the 2 × 2 switch after the cyclability test at 25, 30, and 35 °C when the GST is in the crystalline state as shown in *Fig. S15*. The excellent match among the curves offers good thermal stability.

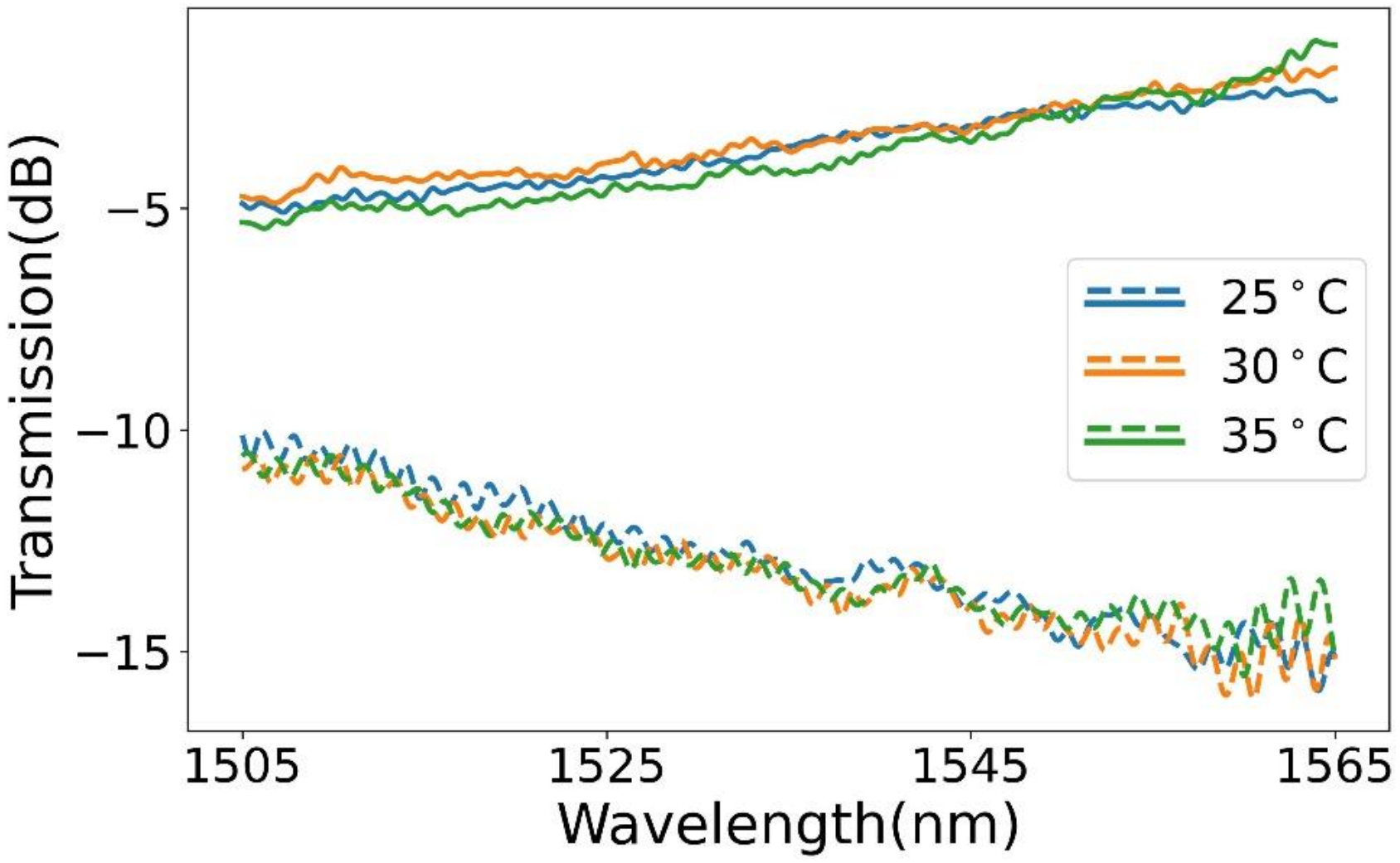

Figure S15. **Thermal stability test measurement.** The temperature varies from 25°C to 35°C. No significant spectrum shift is observed. The cross and bar port transmission are represented by solid and dashed lines, respectively.

# Section 6. SEM images for devices after the endurance test

We show the SEM image of the devices after the cyclability test. In *Fig. S16(a, c)*. The GST thin film shrinks and exhibits a wavey edge. This is likely due to the reflowing(*27*, *35*) while GST was melted or due to non-uniform heating (as can be seen in *Fig. S16(c)*). Some bubbles can also be observed, as in *Fig. S16(b, d)*, which indicates some of the phase change materials are damaged.

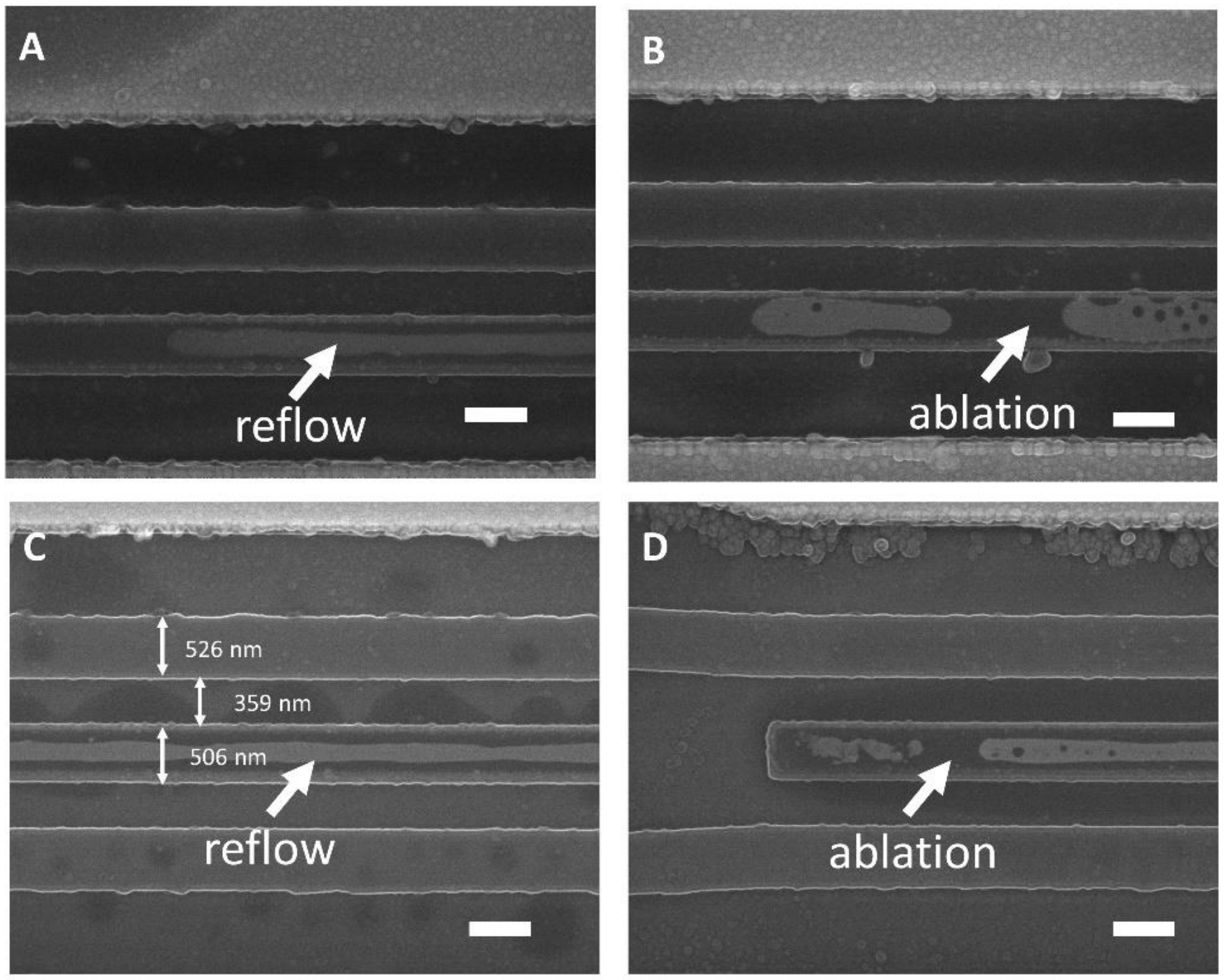

Fig S16. **SEM of the devices after multiple cycles**. (a)(b) 1 × 2 switch reflowing and ablation. (c)(d) 2 × 2 switch reflow and ablation. All scale bars are 500 nm.

# Section 7. System-level applications proposal

The presented nonvolatile 2 × 2 switch, along with the 1 × 2 and 1 × 1 switches, render an energy-efficient and scalable solution to programmable integrated photonics. First, these are nonvolatile devices, and zero static energy is required to hold the state. This is very desirable for low-frequency programmable photonics, as shown in *Table 1*. Second, our device can be switched electrically. System-level and automatic solutions can be envisioned by interfacing with mature electronic technologies. Lastly, the switches are broadband devices, which will enable full use of the parallelism operation of light by wavelength division multiplexing (WDM). Also, the device is relatively insensitive to thermal drifts (Section 5) and fabrication errors because of its broadband nature.

The 2 × 2 switch can be utilized in large-scale nonvolatile optical switching fabric(*36*), as in *Fig. S17(a)*. An 8 × 8 non-blocking optical switching fabric using the classical Benes topology is presented. More applications can be envisioned if the 2 × 2 switch can operate in a multilevel switching fashion, which can be potentially implemented by introducing a detuning between the SW and HW. For example, such devices are essential building blocks for a nonvolatile optical programmable gate array (OPGA). We propose to connect our 2 × 2 devices with each other to form a mesh array, such as in *Fig. S17(b)*. By configuring the optical switch into different states, diverse types of functions can be realized for optical on-chip information processing. Most OFGAs were demonstrated with thermal-optics effect(*6*, *8*, *50*). But due to its volatile nature, a significant amount of energy is consumed to hold the configuration. PCM-based OPGA can significantly save the energy thanks to its self-holding property and thus zero static energy consumption.

Another possible application is optical computing. Recently there have been works on using phase change materials for on-chip non-Von Neumann optical computing(*24*, *51*, *27*). These works mainly apply optical pulses to actuate the phase transition, limiting the device's scalability due to complicated optical alignment. We propose the structure in *Fig. S17(c)*, which can function as an optical forward neural network. Also, similar to (*24*), an optical convolutional neural network comprising 1 × 1 multilevel switches with electrical control can be envisioned, as shown in *Fig. S17(d)*. We expect a large-scale, nonvolatile, and in-memory optical computing platform using electrical controls.

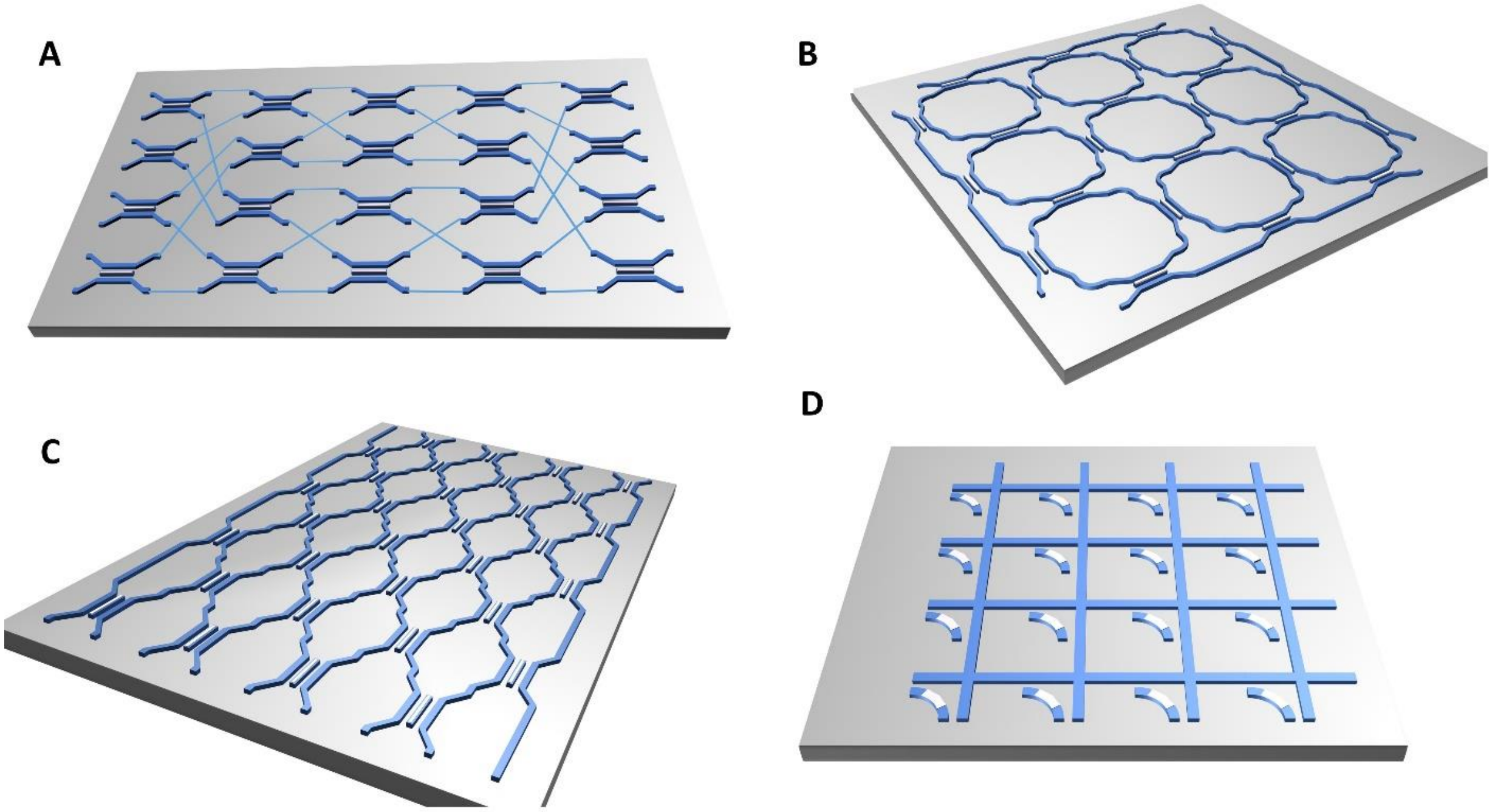

Figure S17. **PIC schematics for various applications.** (a) on-chip optical switching fabric. (b) multipurpose programmable PICs. (c) On-chip optical forward neural networks. (d) convolutional neural networks.